\documentclass[11pt,english]{article}
\pdfoutput=1
\usepackage{jheppub}

\usepackage{amsmath}
\usepackage{amssymb}
\usepackage{amsfonts}
\usepackage{color}

\usepackage{graphicx}

\usepackage{braket}
\usepackage{subfigure}
\usepackage{wrapfig}

\newcommand{\beq}{\begin{equation}}
\newcommand{\eeq}{\end{equation}}
\newcommand{\avg}[1]{{\langle#1\rangle}}

\newcommand{\fft}[2]{\frac{#1}{#2}}
\newcommand{\nn}{{\nonumber}}

\title{Universal features of Lifshitz Green's functions from holography}

\author[a]{Cynthia Keeler,}
\author[b]{Gino Knodel,}
\author[b]{James T. Liu}
\author[b]{and Kai Sun}

\affiliation[a]{Niels Bohr International Academy, Niels Bohr Institute\\
University of Copenhagen,
Blegdamsvej 17, DK 2100, Copenhagen, Denmark}

\affiliation[b]{Michigan Center for Theoretical Physics, Randall Laboratory
of Physics,\\
The University of Michigan, Ann Arbor, MI 48109--1040, USA}

\emailAdd{keeler@nbi.ku.dk}
\emailAdd{gknodel@umich.edu}
\emailAdd{jimliu@umich.edu}
\emailAdd{sunkai@umich.edu}

\abstract{We examine the behavior of the retarded Green's function in theories with Lifshitz scaling
symmetry, both through dual gravitational models and a direct field theory approach.  In contrast
with the case of a relativistic CFT, where the Green's function is fixed (up to normalization) by symmetry,
the generic Lifshitz Green's function can a priori depend on an arbitrary function $\mathcal G(\hat\omega)$,
where $\hat\omega=\omega/|\vec k|^z$ is the scale-invariant ratio of frequency to wavenumber, with
dynamical exponent $z$.  Nevertheless, we demonstrate that the imaginary part of the retarded
Green's function (i.e.\ the spectral function) of scalar operators is exponentially suppressed in a window of frequencies
near zero.  This behavior is universal in all Lifshitz theories without additional constraining symmetries.
On the gravity side, this result is robust against higher derivative corrections, while on the field theory
side we present two $z=2$ examples where the exponential suppression arises from summing
the perturbative expansion to infinite order.}

\begin{document}
\preprint{MCTP-15-10}
\maketitle

\section{Introduction}

The AdS/CFT correspondence provides us with a remarkable strong-weak coupling duality between a bulk gravitational theory and a boundary field theory in one fewer dimension, and as such it 
has found numerous applications in calculating interesting observables
of strongly coupled field theories using weakly coupled holographic
methods.  In recent years,
the number of proposed gravity duals to interesting strongly coupled
QFTs has increased significantly. In addition to the relativistic case
of asymptotically AdS backgrounds, there exist interesting gravitational
duals exhibiting non-relativistic (or Lifshitz-) scaling symmetry \cite{Kachru:2008yh}, as well
as Schr{\"o}dinger symmetry \cite{Son:2008ye,Balasubramanian:2008dm,Adams:2008wt}.  These non-relativistic backgrounds
are particularly relevant for studying condensed matter
systems at strong coupling, where exact analytic results are difficult
to obtain using traditional methods.

In many standard cases, the mapping between bulk and boundary is easily obtained.
This is especially true when there is a brane interpretation, such as the familiar picture
of IIB theory on AdS$_5\times S^5$.  In this case, the AdS$_5$ supergroup
$\mathrm{SU}(2,2|4)$ is identical to the superconformal symmetry group
of the four-dimensional $\mathcal N=4$ super-Yang Mills theory.  As a result, all
observables are constrained by the superconformal symmetry, and in particular
the two-point functions are fully determined up to normalization.  For example,
the retarded scalar Green's function in momentum space must have the form
\begin{equation}
G_R(q^2)=A(-q^2)^{\Delta-2},\qquad q^2=\omega^2-|\vec k\,|^2,
\label{eq:GRrel}
\end{equation}
where $A$ is an overall constant and $\Delta$ is the conformal dimension of the scalar
operator $\mathcal O_\Delta$.

The mapping between condensed matter systems and backgrounds with non-relativistic
scaling symmetry is often less obvious.  In this case, we must often fall back to the general
strategy of constructing a holographic dual to a given field theory by
matching symmetries and conserved quantities \cite{Kachru:2008yh,Son:2008ye,Balasubramanian:2008dm,Adams:2008wt}.  Moreover, non-relativistic
scale invariance is no longer sufficient to fully constrain the form of the two-point
functions.  Consider, for example, the case of Lifshitz scaling with dynamical exponent
$z$, where energy and momentum scale as $\omega\to\lambda^z\omega$ and
$\vec k\to\lambda\vec k$, respectively.  This scaling symmetry only constrains the
form of the Green's function up to an arbitrary function of the scale-invariant quantity
$\hat\omega=\omega/|\vec k\,|^z$:
\begin{equation}
G_R(\omega,\vec k\,)=|\vec k\,|^{2\nu z}\mathcal G(\hat\omega).
\label{eq:GRnr}
\end{equation}
Here $\nu$ is the energy scaling dimension, and
the momentum-dependent prefactor is chosen to give $G_R$ the proper scaling
dimension.

The form of the Green's function (\ref{eq:GRnr}) holds for any (isotropic) scale-invariant
theory, whether computed directly from the field theory or via the holographic dual.  However,
in general, $\mathcal G(\hat\omega)$ cannot be fixed by matching symmetries alone.
(If additional symmetries are imposed, such as $z=2$ Schr\"odinger symmetry, then
the Green's function may become fully determined.)  This suggests that symmetries are
not sufficient for connecting non-relativistic theories to their holographic duals, and in
particular that the duality map must include additional dynamical information.

At the same time, the bulk theory yields a preferred choice of the Green's function obtained from
the classical two-derivative bulk action.  For $z=2$ Lifshitz, the holographic scalar Green's
function was obtained analytically in \cite{Kachru:2008yh}, while a WKB calculation for arbitrary $z>1$
demonstrated a characteristic exponential suppression of the spectral weight (i.e. the imaginary
part of the Green's function) in the limit $\hat\omega\to0$ \cite{Keeler:2014lia}.  This has been interpreted
as an ``insensitivity'' of the boundary theory to small changes
of the geometry near the horizon. 
The same exponential
behavior is responsible for making the smearing function of both Schwarzschild-AdS and Lifshitz spacetime
a distribution rather than a true function \cite{Leichenauer:2013kaa,Keeler:2013msa},
and has been interpreted as a loss of bulk locality for such non-relativistic geometries \cite{Rey:2014dpa}. 

It is natural to expect that different field theoretic models with the same dynamical exponent
$z$ will yield different Green's functions.  This raises the issue as to how the holographic
dual can distinguish among these models.  For unbroken scaling symmetry, the bulk geometry
is essentially fixed to be pure Lifshitz (if we work within the context of general relativity; see e.g. \cite{Hartong:2015zia,Hofman:2014loa,Griffin:2012qx} for other approaches). Thus
the background alone cannot distinguish between different models, and we are mainly left
with the dynamics of the bulk fields as the distinguishing characteristic.  In particular, the
addition of higher derivative terms to the bulk equations of motion will directly affect the form of the
holographic Green's function.  This is in contrast with the relativistic case, where higher
derivative corrections may affect the constant $A$ in (\ref{eq:GRrel}), but will not otherwise
modify the functional form of the retarded Green's function.

Once we allow for a higher derivative expansion in the bulk, it may seem that some predictive
power is lost, since the holographic Green's function would in principle be sensitive to all of the infinitely many
higher derivative terms.  However, we demonstrate that there are universal features that remain.
In particular, the characteristic exponential suppression of the spectral function in the low
frequency regime found in \cite{Keeler:2014lia} is robust with respect to higher derivatives in the bulk, as long as
the frequencies stay above a (momentum-dependent) cutoff.  

We furthermore show that this
exponential suppression arises in field theory models with $z=2$ scaling.  In particular, for both the
quadratic band crossing model of \cite{Sun2009} and the quantum Lifshitz
model \cite{Ardonne2004}, a simple kinematical argument demonstrates that the exponential suppression arises
because one has to go to higher and higher orders in the perturbative expansion to see non-zero
spectral weight in the limit $\hat\omega\to0$.

This paper is organized as follows.  In section~\ref{sec:scaling}, we briefly review the form of the scalar
Green's function in a theory with Lifshitz scaling.  Then in section~\ref{sec:holoLif}, we set up the
computation of the Green's function in a holographic Lifshitz model with bulk higher derivatives.  Following
that, we perform a WKB analysis of the spectral function in section~\ref{sec:WKB} and show
that it has a universal exponential suppression at small $\hat\omega$.  In section~\ref{sec:FT},
we demonstrate that this exponential suppression can be seen directly from a field theory perspective.
We focus on the quadratic band crossing model, but also consider the quantum Lifshitz model.
Finally, we conclude in section~\ref{sec:dis} with a conjecture that this suppression is universal for
all Lifshitz theories in the absence of further constraining symmetries.

\section{The Green's function in a scale invariant theory}
\label{sec:scaling}

In a translationally invariant theory, the retarded Green's function is naturally written in momentum space
as $G_R(\omega,\vec k)$.  Furthermore, unitarity and causality demand that $G_R$ is analytic in
the upper half of the complex $\omega$-plane.  For a scale-invariant theory, the conditions on the Green's
function are much stronger.  In particular, for Lifshitz scaling symmetry with dynamical exponent $z$
\beq\label{eq:Lifscaling}
\vec{x} \rightarrow \Lambda \vec{x}, \qquad t \rightarrow \Lambda^z t,
\eeq
scale and rotational invariance demand that $G_R$ cannot depend on $\omega$ and $\vec k$
separately, but must have the form
\beq\label{eq:scalingGr}
G_R(\omega,\vec k)=|\vec k|^{2\nu z}\mathcal G(\hat\omega)\quad\mbox{where}
\quad\hat\omega\equiv\frac\omega{|\vec k|^z}.
\eeq
Here $\nu$ is the energy scaling dimension and $\mathcal G(\hat\omega)$ is analytic in the upper half
$\hat\omega$ plane.

Non-relativistic scale invariance by itself does not further constrain the form of $\mathcal G(\hat\omega)$.
However, additional symmetries can fix it completely.  For example, relativistic conformal invariance (for the
case $z=1$) constrains $G_R\sim(-q^2)^\nu$ where $q^2=-k_\mu k^\mu=\omega^2-|\vec k|^2$.
This is equivalent to taking the function
\begin{equation}
\mathcal G_{\rm CFT}=A(1-\hat\omega^2)^\nu,
\label{eq:GCFT}
\end{equation}
where $A$ is a constant.  Similarly, full Schr\"odinger symmetry (for $z=2$) \cite{Son:2008ye,Balasubramanian:2008dm,Adams:2008wt} requires 
\begin{equation}
\mathcal G_{\rm Sch}=A(1-2m\hat\omega)^{2\nu},
\label{eq:GSch}
\end{equation}
where $A$ is again a constant, and $m$ is the eigenvalue of the mass-operator of the Schr\"odinger algebra.

While the relativistic and Schr\"odinger cases are the most extensively studied, we are mainly interested
in exploring the features of the function $\mathcal G(\hat\omega)$ for Lifshitz models without additional symmetries using holographic methods.  In general,
$\mathcal G$ will depend on the details of the model.  However, some universal properties can be
deduced in both the small and large $\hat\omega$ limits.  For $\hat\omega\to0$, the only dimensionful
quantity that remains is $|\vec k|$.  Hence $G_R$ must behave as $|\vec k|^{2\nu z}$, or
equivalently
\begin{equation}
\mathcal G(\hat\omega\to0)\sim\mbox{const}.
\label{eq:G0}
\end{equation}
On the other hand, when $\hat\omega\to\infty$, the dependence on $|\vec k|$ drops out, and we must have
\begin{equation}
\mathcal G(\hat\omega\to\infty)\sim\hat\omega^{2\nu}.
\label{eq:Ginfty}
\end{equation}
As can be seen from (\ref{eq:GCFT}) and (\ref{eq:GSch}), the $z=1$ and Schr\"odinger $z=2$ cases both
satisfy these properties.

The retarded Green's function is in general complex, and this ought to be kept in mind when
considering the limiting behaviors given above.  Of particular interest is the general behavior of
the spectral function $\chi(\omega,\vec k)=2\,\mbox{Im}\,G_R(\omega,\vec k)$.  For large $\omega$,
the spectral function scales as $\chi\sim\omega^{2\nu}$, consistent with (\ref{eq:Ginfty}), as well as
the relativistic and Schr\"odinger cases.  The small $\omega$ limit, on the other hand, is more
subtle.  While scaling symmetry demands $\chi\sim2|\vec k|^{2\nu z}
\mbox{\,Im\,}\mathcal G(\hat\omega)$, with $\mbox{\,Im\,}\mathcal G(\hat\omega)$ approaching
a constant as $\hat\omega\to0$, this constant is in fact zero for the $z=1$ and $z=2$ Schr\"odinger
cases.  Moreover, for these cases $\chi$ is identically vanishing for a range of $\hat\omega$ near
zero.  However, this no longer needs to be the case in theories with Lifshitz scaling, but without additional symmetries.
Nevertheless, as we have shown in \cite{Keeler:2014lia}, in the latter case the spectral function is at most exponentially
small in the limit $\hat\omega\to0$, at least in the two-derivative holographic theory.  What we
will show below is that this exponential suppression of $\chi$ remains robust, even when higher
derivative corrections are included, as long as the perturbative expansion is kept under control.

\section{Holographic Lifshitz models}
\label{sec:holoLif}

The Lifshitz symmetry (\ref{eq:Lifscaling}) can be realized in a gravitational background given by the
metric
\begin{equation}
ds_{d+2}^2=\fft{-dt^2+d\rho^2}{\rho^2}+\fft{d\vec x^2}{\rho^{2/z}}.
\label{eq:asympmetric}
\end{equation}
The boundary of the bulk spacetime is located at $\rho=0$, while the horizon is at $\rho=\infty$.
For simplicity, we examine the scalar Green's function, which can be holographically computed from
the action of a bulk scalar $\phi(t,\vec x,\rho)$.

At the two-derivative level, the minimally coupled equation of motion for $\phi$ is simply
$(\square-m^2)\phi=0$.  This system has been extensively studied, and the holographic computation of
the retarded Green's function is by now standard \cite{Son:2002sd}.  Working in momentum space and taking
\beq\label{eq:phidef}
\phi(t,\vec x,\rho) = e^{i(\vec{k}\cdot\vec{x}-\omega t)}\rho^{d/2z} \psi (\rho),
\eeq
we find that $\psi(\rho)$ satisfies the Schr\"odinger-like equation $-\psi''+U_0\psi=0$ where
\begin{equation}
U_0= \frac{\nu^2-1/4}{\rho^2}+\frac{|\vec k|^2}{\rho^{2-2/z}}-\omega^2.
\end{equation}
and
\begin{equation}
\nu=\sqrt{m^2+\left(\fft{d+z}{2z}\right)^2}.
\label{eq:nufromm}
\end{equation}
We can highlight the scaling properties of the solution by defining the dimensionless coordinate
\begin{equation}
\hat\rho=\rho|\vec k|^z. \label{eq:rhohat_def}
\end{equation}
The Schr\"odinger-like equation now takes the form
\begin{equation}
-\psi''(\hat\rho)+\hat U_0(\hat\rho)\psi(\hat\rho)=0,\qquad
\hat U_0(\hat\rho)= \frac{\nu^2-1/4}{\hat\rho^2}+\frac{1}{\hat\rho^{2-2/z}}-\hat\omega^2.
\label{eq:Slehat}
\end{equation}
In order to apply the AdS/CFT prescription for calculating the retarded Green's function, we need to examine the
solution near the boundary at $\hat\rho=0$ and as it approaches the horizon at $\hat\rho=\infty$.
In the limit $\hat\rho\to0$, the Schr\"odinger potential
is dominated by the $(\nu^2-1/4)/\hat\rho^2$ term, and we find the boundary behavior
\begin{equation}
\psi(\hat\rho\to0)\sim A\hat\rho^{\fft12-\nu}+B\hat\rho^{\fft12+\nu}.
\label{eq:bdyAB}
\end{equation}
Here we have used the convention that $B$ is the coefficient of the normalizable mode, while $A$ is
the coefficient of the non-normalizable mode.  For $z>1$, $\hat U_0$ approaches $-\hat\omega^2$ at
the horizon, so the solution is oscillatory:
\begin{equation}
\psi(\hat\rho\to\infty)\sim ae^{i\hat\omega\hat\rho}+be^{-i\hat\omega\hat\rho}.
\end{equation}
For the retarded Green's function, we take infalling boundary conditions, which correspond to setting
$b=0$.  In this case, we find
\begin{equation}
\mathcal G(\hat\omega)=\left.\fft{B}A\right|_{b=0},
\label{eq:G=B/A}
\end{equation}
where the relation between $\{A,B\}$ at the boundary and $\{a,b\}$ at the horizon is obtained by
solving the Schr\"odinger problem (\ref{eq:Slehat}).

\subsection{Bulk higher derivatives}

At the two-derivative level, the solution for $\mathcal G(\hat\omega)$ has been extensively studied,
and analytic results
may be obtained for $z=1$ and $z=2$ \cite{Son:2002sd,Kachru:2008yh}.  However, as we emphasized in section \ref{sec:scaling}, scaling symmetry by itself does not fully constrain
the form of the Green's function.  This raises the question of where the freedom of arbitrarily choosing
the function $\mathcal G$ arises in the holographic dual.  If we work within general relativity, there
are two natural possibilities:
the first is the choice of background metric, and the second is the form of the scalar equation.
However, the metric (\ref{eq:asympmetric}) is essentially unique
(up to coordinate transformations) once we have imposed Lifshitz scaling.  This leaves us with modification
of the equation of motion.

From a bulk effective field theory point of view, it is possible to include higher derivative terms
in the scalar equation.  In momentum space, non-radial derivatives in the effective action show up as
powers of $\omega$ and $\vec k$, while additional $\rho$ derivatives lead to a higher order
differential equation for $\psi(\rho)$.  If there are no additional $\rho$ derivatives, then the momentum
space equation remains second order and can be brought into Schr\"odinger form just as above.
This time, however, the effective Schr\"odinger potential in (\ref{eq:Slehat}) generalizes to
\beq\label{eq:Udef}
\hat U(\hat\rho)= \frac{\nu^2-1/4}{\hat\rho^2}+\fft1{\hat\rho^{2-2/z}}-\hat\omega^2
+\frac{1}{\hat\rho^2}{f} (\hat\omega\hat\rho,\hat\rho^{1/z}).
\eeq
where the function $f$ encodes the presence of the higher derivative terms. 

In principle, the procedure for extracting the holographic Green's function is unchanged from the prescription of
(\ref{eq:G=B/A}).  However, the higher derivative terms affect the shape of the potential, and
hence may change the boundary and horizon asymptotics and possibly also introduce
additional classical turning points in the bulk.  In order to get a better understanding of the asymptotics,
we write out the expansion
\begin{equation}
\fft1{\hat\rho^2}f(\hat\omega\hat\rho,\hat\rho^{1/z})=\sum_{\substack{i,j\\i+j>2}}\lambda_{i,j}\hat\omega^i \hat\rho^{i+j/z-2},
\label{eq:fexp}
\end{equation}
where $i$ and $j$ count the number of temporal and spatial derivatives, respectively.  
The restriction $i+j>2$ ensures that $f$ only comprises the higher derivative contributions.
Note that the coefficients
$\lambda_{i,j}$ are dimensionless, although (after restoring units) we typically expect $\lambda_{i,j}\sim(\ell/L)^{i+j-2}$,
where $\ell$ is some microscopic scale and $L$ is the curvature scale of the Lifshitz bulk, such
that $\ell\ll L$.

Focusing first on the boundary at $\hat\rho=0$, we see that the behavior of the potential
(\ref{eq:Udef}) remains dominated by the $1/\hat\rho^2$ term, since $i+j>2$ in the derivative expansion.  Thus the boundary scaling behavior remains unchanged from (\ref{eq:bdyAB}),
and the relation of the scaling dimension to $\nu$ is unaffected by the higher order terms.

The horizon behavior, on the other hand, is considerably different.  Since the horizon is located at
$\hat\rho\to\infty$, and the expansion (\ref{eq:fexp}) in general contains positive powers of $\hat\rho$,
the successive higher derivative terms will become more and more dominant at the horizon.
Furthermore, the potential will generically go to $\pm \infty$ at the horizon,
depending on the sign of $\lambda_{i,j}$ of the dominant term.  As a result, strictly speaking,
the perturbative expansion of the scalar equation breaks down near the horizon.  Nevertheless,
we now argue that the holographic Green's function can be extracted from the solution of the higher
derivative equation in a controlled manner.

\subsection{Consistency of the higher derivative expansion}
\label{sec:consistency}

At the two-derivative level, the Schr\"odinger potential (\ref{eq:Slehat}) is monotonically decreasing as
we move into the interior of the bulk, and there is a single classical turning point located at $\hat\rho_0$
where $\hat U_0(\rho_0)=0$.  For $\hat\rho<\hat\rho_0$, the solution connects to the power-law behavior
(\ref{eq:bdyAB}) at the boundary, while for $\hat\rho>\hat\rho_0$, the solution is oscillatory, and
infalling boundary conditions are chosen at the horizon.

Ignoring the shift of $\nu^2$ in (\ref{eq:Slehat}), there are two competing power laws in $\hat U$,
namely $\nu^2/\hat\rho^2$ and $1/\hat\rho^{2-2/z}$, and the behavior of the solution depends on which
of the power laws dominates at the classical turning point.  We define the crossover point as
$\hat\rho_*=\nu^z$, which is the location where the two terms become comparable.  There are
two distinct cases to consider:
\begin{enumerate}
\item For $\hat\omega\gg\nu^{1-z}$, the classical turning point is located at $\hat\rho_0\approx
\nu/\hat\omega\ll\hat\rho_*$.  This point is close to the boundary, and the $1/\hat\rho^2$ potential
ensures a power law behavior without exponential suppression.  The holographic Green's function
is ``featureless'', and behaves as $\mathcal G\sim\hat\omega^{2\nu}$.
\item For $\hat\omega\ll\nu^{1-z}$, the classical turning point is instead located at
$\hat\rho_0\approx\omega^{-z/(z-1)}\gg\hat\rho_*$.  The Green's function now probes deep into
the bulk, and can have non-trivial features.  Note that the wavefunction has
exponential behavior in the region $\hat\rho_*<\hat\rho<\hat\rho_0$, leading to an effective
decoupling of the boundary from the horizon \cite{Keeler:2014lia}.
\end{enumerate}
We now consider the effect of the higher derivative terms, encoded in the function $f$ in
(\ref{eq:fexp}).  Although this function dominates at the horizon, we nevertheless consider
a formal perturbative expansion of the Schr\"odinger problem in the couplings $\lambda_{i,j}$.
Of course, the higher order terms will dominate the wavefunction near the horizon.  However, it
is important to realize that the holographic Green's function is not determined by the wavefunction
at the horizon, but by its asymptotic behavior at the boundary.  Infalling boundary conditions are
needed at the horizon, but this can be imposed consistently at each order in the perturbative
expansion.  These infalling conditions will be seen in the boundary Green's function, but will
not dominate over lower orders in the expansion.

Although a formal perturbative expansion can be used to solve the bulk scalar equation, the
expansion of $\mathcal G$ in the couplings $\lambda_{i,j}$ will
only be sensible if the corrections can be kept small.  Obviously this cannot be true globally, as
the higher derivative terms typically dominate near the horizon.  However, as one can see for example by using the WKB approximation (see section \ref{sec:WKB} and \cite{Keeler:2014lia}), the holographic
Green's function only gives us information about physics between the boundary and the classical turning point $\hat\rho_0$, where the wavefunction changes from exponential to oscillating behavior.
Hence all that is necessary is to ensure that $f$ remains small compared to the leading order potential
$\hat U_0$ only for $\hat{\rho}\leq \hat{\rho}_0$.  The specifics of this condition depend on whether we are in the high or low frequency
regime.  We consider these two cases separately:
\begin{enumerate}
\item In the high frequency regime ($\hat\omega\gg\nu^{1-z}$), the dominant term in $\hat U_0$ is
$\nu^2/\hat\rho^2$.  Since this term is decaying, while at the same time $f$ becomes more important
as we move away from the boundary, we only need to demand that $f$ is small compared to
$\nu^2/\hat\rho^2$ at the classical turning point.  This gives rise to the condition
$f(\hat\omega\hat\rho_0,\hat\rho_0^{1/z})\ll\nu^2$, which may be satisfied by taking $(\ell/L)\nu\ll1$,
where we have assumed the expansion (\ref{eq:fexp}) along with the behavior of the couplings
$\lambda_{i,j}\sim(\ell/L)^{i+j-2}$.  As we may see from (\ref{eq:nufromm}), the scale of $\nu$
is set by $mL$.  Therefore, the condition for a valid expansion is equivalent to demanding $m\ell\ll 1$.
We conclude that in this case, higher derivative corrections are under perturbative control provided the
bulk couplings satisfy $m\ell\ll 1$.  This behavior is very much like the
relativistic $z=1$ case, since in both situations the $\nu^2/\hat\rho^2$ potential dominates up to the
classical turning point.  
\item In the low frequency regime ($\hat\omega\ll\nu^{1-z}$), we need to compare $f$ with
the $1/\hat\rho^{2-2/z}$ term in $\hat U_0$.  Once again, we only need to consider the magnitude
of $f$ at the classical turning point.  The condition is now
$f(\hat\omega\hat\rho_0,\hat\rho_0^{1/z})\ll\hat\rho_0^{2/z}$, which gives rise to the requirement
\begin{equation}
\hat\omega\gg\left(\fft\ell{L}\right)^{z-1}.
\label{eq:wmin}
\end{equation}
As $\hat\omega$ is taken smaller and smaller, we need to take higher and higher order corrections
into account.  As a result, the perturbative expansion breaks down at small $\hat\omega$, and results
computed in this regime will not be robust against higher derivative corrections.  Physically, what happens
is that as $\hat\omega\to0$, we probe closer and closer to the horizon, and it is precisely there where
the higher derivative corrections dominate.
\end{enumerate}
Hence, as long as the scale of the bulk higher derivative corrections
satisfies $m\ell\ll 1$, the perturbative expansion of the boundary Green's function makes sense
for dimensionless frequencies $\hat\omega\gg(\ell/L)^{z-1}$.  For lower frequencies, the
higher derivative terms start dominating.  

This feature of higher derivative terms becoming more pronounced at the horizon is not
restricted to the Lifshitz background, but is in fact fairly general and shows up in,
{\it e.g.}, the pure AdS and Schwarzschild-AdS cases.  While the pure AdS case tends to be
robust against higher derivatives because of conformal invariance, more care
may be needed in the case of holography at non-zero temperature \cite{Policastro:2001yc,Policastro:2002se,Policastro:2002tn,Son:2002sd,Herzog:2002fn,Herzog:2003ke,Kovtun:2004de,Buchel:2004di}.
Transport coefficients, such as the shear viscosity, may be extracted using the Kubo formula, which
is evaluated at $|\vec{k}|=0$ before sending $\omega\to0$.  Since this is consistent with
(\ref{eq:wmin}), the perturbative expansion for transport coefficients is valid.  At the same time,
however, more care may be needed when analyzing general hydrodynamic modes, which are
defined for both $\omega$ and $\vec{k}$ small, but nonzero (see e.g.\ \cite{Son:2007vk}).

\section{WKB analysis of the spectral function}
\label{sec:WKB}

In this section, we study the holographic spectral function of a probe
scalar in Lifshitz spacetime, in the presence of higher derivative
corrections. To determine the effect of higher derivatives on the
retarded Green's function, we consider a probe scalar with an effective
potential of the form 
\begin{equation}
\hat{U}=\frac{\nu^{2}-1/4}{\hat{\rho}^{2}}+\frac{1}{\hat{\rho}^{2-2/z}}-\hat{\omega}^{2}+\sum_{i+j>2}\lambda_{i,j}\hat{\omega}^{i}\hat{\rho}^{i+j/z-2}.\label{eq:U_WKB}
\end{equation}
The last term encodes an infinite set of higher derivative
corrections to the equation of motion, where the $(i,j)$ term corresponds
to $i$ temporal and $j$ spatial derivatives. The size of the coefficients is expected to be set by a microscopic length scale $\ell$, so that (after restoring units of $L$) $\lambda_{i,j}\sim (\ell/L)^{i+j-2}$.
Since it is in general
not possible to solve the corresponding Schr{\"o}dinger equation for the potential (\ref{eq:U_WKB}) analytically,
we will make use of the WKB approximation to obtain an approximate
solution. This method can be used to calculate the imaginary part
of the retarded Green's function, which is proportional to the spectral
function. After switching to the $\hat{\rho}$ coordinates defined
in (\ref{eq:rhohat_def}), the spectral function can be approximated
by%
\footnote{The additional prefactor of $|\vec{k}|^{2\nu z}$ arises from letting
$\epsilon\rightarrow|\vec{k}|^{-z}\epsilon$, which is the proper
UV cutoff needed to cancel the log-divergence of the integral. %
} \cite{Keeler:2014lia} 
\begin{equation}
K^{-1}\mathrm{Im}\,G_{R}(\omega,\vec{k})\approx|\vec{k}|^{2\nu z}\lim_{\epsilon\rightarrow0}\epsilon^{-2\nu}e^{-2S}.\label{eq:ImG_WKB}
\end{equation}
Here $K$ is a normalization constant and 
\begin{equation}
S=\int_{\epsilon}^{\hat{\rho}_{0}}d\hat{\rho}\sqrt{\hat{U}(\hat{\rho})+\frac{1}{4\hat{\rho}^{2}}}.\label{eq:S_WKB}
\end{equation}
The additional $1/\hat{\rho}^{2}$ term is equivalent to an effective
shift $\text{\ensuremath{\nu}}^{2}\rightarrow\nu^{2}+\frac{1}{4}$,
which is necessary for consistency of the WKB approximation for $1/x^{2}$
 potentials \cite{Keeler:2013msa}. The integral is taken from a UV
cutoff $\epsilon$ to the classical turning point $\hat{\rho}_{0}$.
The WKB approximation for the imaginary
part of the rescaled Green's function defined in (\ref{eq:scalingGr})
is given by 
\begin{equation}
K^{-1}\mathrm{Im}\,{\cal G}(\hat{\omega})\approx\lim_{\epsilon\rightarrow0}\epsilon^{-2\nu}e^{-2S}.\label{eq:ImCurlyG_WKB}
\end{equation}
This expression is valid for a potential with only one classical turning
point, such that the wavefunction is oscillating near the horizon
and tunnels towards the boundary. Close to the boundary, the $1/\hat{\rho}^{2}$ part
of the potential leads to a power-law scaling of the wavefunction,
which is stripped off by the factor of $\epsilon^{-2\nu}$ in (\ref{eq:ImCurlyG_WKB}).
We can use (\ref{eq:ImCurlyG_WKB}) to determine the imprint of higher
derivative corrections on the spectral function, provided that $\lambda_{i,j}<0$.
In this case, the potential goes to $-\infty$ at the horizon, but
the wavefunction still remains oscillating and we can consistently
impose infalling boundary conditions. Later we will argue that (\ref{eq:ImCurlyG_WKB})
can in fact be used to provide a formal expansion for corrections
with arbitrary sign.

In order to perform a perturbative expansion of the WKB integral (\ref{eq:S_WKB})
in terms of $\lambda_{i,j}$, we need the higher derivative corrections
to be subdominant compared to the other terms in $\hat{U}$, at least
in the domain of integration. We therefore demand 
\begin{equation}
\lambda_{i,j}\hat{\omega}^{i}\hat{\rho}^{i+j/z}\ll\nu^{2},\qquad\lambda_{i,j}\hat{\omega}^{i}\hat{\rho}^{i+j/z}\ll\hat{\rho}^{2/z}\label{eq:smallness}
\end{equation}
for all $0<\hat{\rho}\leq\hat{\rho}_{0}$ (see also the discussion in section
\ref{sec:consistency}). We can already see that this imposes an $\hat{\omega}$-dependent
condition on the coefficients $\lambda_{i,j}$, which we will make
more explicit in what follows.

We can now determine the leading order correction to $\mathrm{Im}\,{\cal G}(\hat{\omega})$
by formally expanding the WKB integral in terms of the $\lambda_{i,j}$.
At leading order, the higher derivative contributions are linear,
so for our purposes it will be enough to drop the sum in (\ref{eq:U_WKB}) and only
consider the effect of a single correction term with fixed $(i,j)$.
In a realistic model with a tower of higher derivative corrections,
one may obtain a perturbative expansion for $\mathrm{Im}\,{\cal G}(\hat{\omega})$
by summing up the individual contributions, keeping in mind that if
there are corrections at different order (e.g. $\alpha^{\prime}$ and
$\left(\alpha^{\prime}\right)^{2}$), one may have to go beyond linear
order to study the effect of all correction terms.

A consistent expansion in $\lambda_{i,j}$ requires expanding both
the integrand and the upper bound $\hat{\rho}_{0}$, since the location
of the turning point depends on the details of the correction terms.
Writing $S=S^{(0)}+\delta S$, where $S^{(0)}$ is the two-derivative
integral with $\lambda_{i,j}=0$, we find (see appendix \ref{app:pewkb} for a rigorous
derivation): 
\begin{align}
S^{(0)} & =\int_{\epsilon}^{\hat{\rho}_{0}^{(0)}}d\hat{\rho}\sqrt{\frac{\nu^{2}}{\hat{\rho}^{2}}\text{+}\frac{1}{\hat{\rho}^{2-2/z}}-\hat{\omega}^{2}},\label{eq:S_0}\\
\delta S & \approx\int_{\epsilon}^{\hat{\rho}_{0}^{(0)}}d\hat{\rho}\frac{\lambda_{i,j}\hat{\omega}^{i}\hat{\rho}^{i+j/z-2}}{2\sqrt{\frac{\nu^{2}}{\hat{\rho}^{2}}\text{+}\frac{1}{\hat{\rho}^{2-2/z}}-\hat{\omega}^{2}}},\label{eq:delta_S}
\end{align}
where $\hat{\rho}_{0}^{(0)}$ is the turning point for the case $\lambda_{i,j}=0$,
i.e. the solution of 
\begin{equation}
\frac{\nu^{2}}{\hat{\rho}_{0}^{2}}\text{+}\frac{1}{\hat{\rho}_{0}^{2-2/z}}-\hat{\omega}^{2}=0,
\end{equation}
and we expanded up to linear order in $\lambda_{i,j}$. The large
and small $\hat{\omega}$-behavior of the unperturbed integral $S^{(0)}$
was computed in \cite{Keeler:2013msa}: 
\begin{align}
S^{(0)}(\hat{\omega}\gg\nu^{1-z}) & \approx-\nu-\nu\log\left(\frac{\epsilon}{2\nu}\right),\label{eq:S0_largef}\\
S^{(0)}(\hat{\omega}\ll\nu^{1-z}) & \approx-z\nu+z\nu\log\left(2\nu\right)+\nu(z-1)\log z+\frac{\sqrt{\pi}\Gamma\left(\frac{1}{2(z-1)}\right)}{z\Gamma\left(\frac{z}{2(z-1)}\right)}\hat{\omega}^{-\frac{1}{z-1}}.\label{eq:S0_smallf}
\end{align}
Let us now calculate the leading correction (\ref{eq:delta_S}) in
the same limits. For $\hat{\omega}\gg\nu^{1-z}$, the unperturbed
turning point lies at $\hat{\rho}_{0}^{(0)}\approx \nu/\hat{\omega}$,
which is well within the region where the $1/\hat{\rho}^{2}$ term
dominates over $1/\hat{\rho}^{2-2/z}$. Hence we can approximate the
integral as 
\begin{equation}
\delta S\approx\int_{\epsilon}^{\nu/\hat{\omega}}d\hat{\rho}\frac{\lambda_{i,j}\hat{\omega}^{i}\hat{\rho}^{i+j/z-2}}{2\sqrt{\frac{\nu^{2}}{\hat{\rho}^{2}}-\hat{\omega}^{2}}}.\label{eq:delta_S_largeOmega}
\end{equation}
Letting $x\equiv\hat{\omega}\hat{\rho}/\nu$, we find 
\begin{eqnarray}
\delta S & \approx & \nu\lambda_{i,j}\nu^{i+j-2}\left(\frac{\nu^{1-z}}{\hat{\omega}}\right)^{\frac{j}{z}}\int_{\frac{\epsilon\hat{\omega}}{\nu}}^{1}dx\frac{x^{i+\frac{j}{z}-2}}{2\sqrt{\frac{1}{x^{2}}-1}}.
\end{eqnarray}
For $\hat{\omega}\gg\nu^{1-z}$, correction terms with $j\neq0$ are
highly suppressed. After taking the UV cutoff $\epsilon$ to zero,
we therefore have 
\begin{equation}
\delta S\approx\delta_{j,0}c_{i}\lambda_{i,0}\nu^{i-1}+O\left(\frac{\nu^{1/z-1}}{\hat{\omega}^{1/z}}\right),\label{eq:delta_S_result1}
\end{equation}
where 
\begin{equation}
c_{i}=\int_{0}^{1}dx\frac{x^{i-1}}{2\sqrt{1-x^{2}}}=\frac{\sqrt{\pi}\Gamma\left(\frac{i}{2}\right)}{4\Gamma\left(\frac{i+1}{2}\right)}.
\end{equation}
Using (\ref{eq:delta_S_result1}) and the unperturbed result (\ref{eq:S0_largef}),
we arrive at the final answer 
\begin{align}
K^{-1}\mathrm{Im}\,{\cal G}(\hat{\omega}\gg\nu^{1-z}) & \approx C\hat{\omega}^{2\nu},\qquad C=(2\nu)^{-2\nu}\exp\left[2\nu\left(1-\delta_{j,0}c_{i}\lambda_{i,0}\nu^{i-2}+\cdots\right)\right],\label{eq:ImG_largeomega}
\end{align}
where the ellipsis indicates terms that are higher order in $\lambda$.
The scaling of ${\cal G}$ with $\hat{\omega}^{2\nu}$ reflects the
fact that at large frequencies, the Green's function $G_{R}(\omega,\vec{k})=|\vec{k}|^{2\nu z}{\cal G}(\hat{\omega})$
becomes independent of $\vec{k}$ (see the discussion in section \ref{sec:scaling}).
The higher derivatives simply renormalize the numerical prefactor
in a controlled way. The size of the higher derivative corrections
at large $\hat{\omega}$ is controlled by 
\begin{equation}
\lambda_{i,0}\nu^{i-2}\sim\left(\frac{\ell}{L}\nu\right)^{i-2}\sim\left(m\ell\right)^{i-2},\label{eq:lambda_nu}
\end{equation}
where $i>2$ is the number of temporal derivatives. Note that $\lambda_{i,0}\nu^{i-2}\ll1$
is precisely what is required for the higher derivative corrections
to be small up to the classical turning point $\hat{\rho}_{0}$, in
the limit of large $\hat{\omega}$, as one can see by evaluating (\ref{eq:smallness})
at $\hat{\rho}_{0}$ in this limit, and noting that the unperturbed
potential is monotonically decreasing.

We now turn to calculating the higher derivative corrections in the
case of small frequencies ($\hat{\omega}\ll\nu^{1-z}$). In this case,
the unperturbed classical turning point lies at $\hat{\rho}_{0}^{(0)}\approx\hat{\omega}^{-z/(z-1)}$.
We can split up the integral (\ref{eq:delta_S}) in the following
way \cite{Keeler:2013msa}: Let $\hat{\rho}_{\ast}=\nu^{z}$ be the
crossover scale, defined in the beginning of section \ref{sec:consistency},
at which the two different terms in the potential (\ref{eq:Udef}),
$\nu^{2}/\hat{\rho}^{2}$ and $1/\hat{\rho}^{2-2/z}$, become comparable.
Since $\hat{\omega}\ll\nu^{1-z}$, we can then introduce a regulator
scale $\hat{\rho}_{r}$ such that $\hat{\rho}_{\ast}\ll\hat{\rho}_{r}\ll\hat{\rho}_{0}^{(0)}$,
and split up the WKB integral in (\ref{eq:delta_S}) as 
\begin{equation}
\int_{\epsilon}^{\hat{\rho}_{0}^{(0)}}=\int_{\epsilon}^{\hat{\rho}_{r}}+\int_{\hat{\rho}_{r}}^{\hat{\rho}_{0}^{(0)}}.
\end{equation}
The first of the integrals above is taken over $\epsilon\leq\hat{\rho}\leq\hat{\rho}_{r}\ll\hat{\rho}_{0}^{(0)}$,
so we can approximate the potential in this region as 
\begin{equation}
\hat{U}\approx\frac{\nu^{2}}{\hat{\rho}^{2}}\text{+}\frac{1}{\hat{\rho}^{2-2/z}}+\lambda_{i,j}\hat{\omega}^{i}\hat{\rho}^{i+j/z-2}.
\end{equation}
On the other hand, the second integral is taken over $\hat{\rho}_{r}\leq\hat{\rho}\leq\hat{\rho}_{0}^{(0)}$,
so in this region we can write 
\begin{equation}
\hat{U}\approx\frac{1}{\hat{\rho}^{2-2/z}}-\hat{\omega}^{2}+\lambda_{i,j}\hat{\omega}^{i}\hat{\rho}^{i+j/z-2}.
\end{equation}
Using these approximations, we find 
\begin{align}
\delta S & =\delta S_{1}+\delta S_{2},\nn\\
 & \approx\int_{\epsilon}^{\hat{\rho}_{r}}d\hat{\rho}\frac{\lambda_{i,j}\hat{\omega}^{i}\hat{\rho}^{i+j/z-2}}{2\sqrt{\frac{\nu^{2}}{\hat{\rho}^{2}}\text{+}\frac{1}{\hat{\rho}^{2-2/z}}}}+\int_{\hat{\rho}_{r}}^{\hat{\rho}_{0}^{(0)}}d\hat{\rho}\frac{\lambda_{i,j}\hat{\omega}^{i}\hat{\rho}^{i+j/z-2}}{2\sqrt{\frac{1}{\hat{\rho}^{2-2/z}}-\hat{\omega}^{2}}}.
\end{align}
Letting $u=\frac{1}{\nu^{2}}\hat{\rho}^{2/z}$, the first
integral can be written as 
\begin{equation}
\delta S_{1}=\frac{z\nu}{4}\lambda_{i,j}\nu^{i+j-2}\left(\frac{\hat{\omega}}{\nu^{1-z}}\right)^{i}\int_{u_{\epsilon}}^{u_{r}}du\frac{u^{\frac{z}{2}i+\frac{j}{2}-1}}{\sqrt{1+u}},
\end{equation}
where the integration bounds are $u_{\epsilon}=\epsilon^{2/z}/\nu^{2}\rightarrow0$
and $u_{r}=\left(\hat{\rho}_{r}/\hat{\rho}_{\ast}\right)^{2/z}\gg1$.
In the small frequency limit $\hat{\omega}\ll\nu^{1-z}$, the correction
term is highly suppressed unless $i=0$. Hence we have 
\begin{equation}
\delta S_{1}\approx\delta_{i,0}\frac{z\nu}{4}\lambda_{0,j}\nu^{j-2}\int_{u_{\epsilon}}^{u_{r}}du\frac{u^{\frac{j}{2}-1}}{\sqrt{1+u}}+O\left(\frac{\hat{\omega}}{\nu^{1-z}}\right).\label{eq:delta_S1}
\end{equation}
The remaining integral is divergent as $u_{r}\rightarrow\infty$.
However, one can show that the contribution of the upper bound cancels
with that from the lower bound of $\delta S_{2}$, since $u_{r}$
is after all a fictitious regulator scale. Hence the only contribution
of (\ref{eq:delta_S1}) to $\delta S$ is due to evaluating the integral
at the lower bound $u_{\epsilon}\rightarrow0$: 
\begin{equation}
\delta S_{1}\rightarrow-\delta_{i,0}\frac{z\nu}{4}d_{j}\lambda_{0,j}\nu^{j-2}+O\left(\frac{\hat{\omega}}{\nu^{1-z}}\right),\label{eq:delta_S1_result}
\end{equation}
where 
\begin{equation}
d_{j}=\int^{u=0}du\frac{u^{\frac{j}{2}-1}}{\sqrt{1+u}}.
\end{equation}
This contribution is finite; in particular there are no $\mathrm{log}\,\epsilon$ terms,
which would affect the boundary scaling. Similar to the large $\hat{\omega}$
case, higher derivative corrections are controlled by terms of order
$\sim\lambda\nu^{n-2}$, where $n$ counts the number of derivatives.
This becomes qualitatively different when considering $\delta S_{2}$,
which captures the contribution of higher derivative corrections
deep in the bulk. Letting $x=\hat{\omega}^{z/(z-1)}\hat{\rho}$,
we obtain 
\begin{equation}
\delta S_{2}=\frac{1}{2}\hat{\omega}^{-\frac{1}{z-1}}e_{i,j}\lambda_{i,j}\hat{\omega}^{-\frac{1}{z-1}(i+j-2)},\label{eq:delta_S2_result}
\end{equation}
where 
\begin{equation}
e_{i,j}=\int_{{\hat{\rho}_{r}}/{\hat{\rho}_{0}^{(0)}}}^{1}dx\frac{x^{i-1+\frac{j-1}{z}}}{\sqrt{1-x^{2-\frac{2}{z}}}}.
\end{equation}
When expanding the lower bound in powers of $\hat{\rho}_{r}/\hat{\rho}_{0}^{(0)}$,
each term is designed to cancel with the corresponding contribution
from $\delta S_{1}$. Instead of carrying out this cancellation explicitly,
we can therefore let 
\begin{equation}
e_{i,j}\rightarrow\int_{0}^{1}dx\frac{x^{i-1+\frac{j-1}{z}}}{\sqrt{1-x^{2-\frac{2}{z}}}}=\frac{\sqrt{\pi}\Gamma\left(\frac{iz+j-1}{2(z-1)}\right)}{(2-\frac{2}{z})\Gamma\left(\frac{(i+1)z+j-2}{2(z-1)}\right)},
\end{equation}
together with the prescription (\ref{eq:delta_S1_result}). Using
(\ref{eq:delta_S1_result}), (\ref{eq:delta_S2_result}) and the zeroth
order result (\ref{eq:S0_smallf}), we arrive at the final answer
\begin{equation}
K^{-1}\mathrm{Im}\,{\cal G}(\hat{\omega}\ll\nu^{1-z})\approx D\exp\left[-\hat{\omega}^{-\frac{1}{z-1}}E(\hat{\omega})\right],\label{eq:ImG_smallOmega}
\end{equation}
where 
\begin{align}
D & =(2\nu)^{-2z\nu}z^{2\nu(1-z)}\exp\bigg[2z\nu\big(1+\delta_{i,0}d_{j}\lambda_{0,j}\nu^{j-2}+\cdots\big)\bigg],\nn\\
E(\hat{\omega}) & =\frac{\sqrt{\pi}\Gamma\left(\frac{1}{2(z-1)}\right)}{z\Gamma\left(\frac{z}{2(z-1)}\right)}+e_{i,j}\lambda_{i,j}\hat{\omega}^{-\frac{1}{z-1}(i+j-2)}+\cdots.
\end{align}
Here the ellipses indicate terms that are higher order in $\lambda$.
We see that the higher derivative terms have two distinct effects:
First, corrections with $i=0$, which correspond to purely spatial
derivatives, affect the overall normalization $D$ of the spectral
function. Second, and more importantly, higher derivative corrections
with any $i$ and $j$ change the behavior of the spectral function
as $\hat{\omega}\rightarrow0$, encoded in $E(\hat{\omega})$. The
$\hat{\omega}$-dependent correction terms become more and more important
at small frequencies, and eventually the perturbative expansion breaks
down. This was to be expected, since at small $\hat{\omega}$, the
spectral function probes deep into the bulk, where higher derivatives
dominate. However, recall that the coupling constants $\lambda_{i,j}$
are generically given by a ratio of a microscopic versus macroscopic length
scale, $\lambda_{i,j}\sim(\ell/L)^{i+j-2}$. It is thus possible to
keep the corrections in (\ref{eq:ImG_smallOmega}) small by demanding

\begin{equation}
\nu^{1-z}\gg\hat{\omega}\gg\left(\frac{\ell}{L}\right)^{z-1}.\label{eq:omega_range}
\end{equation}
This is precisely the bound we argued for in section \ref{sec:consistency}.
Since the condition (\ref{eq:smallness}), evaluated at large $\hat{\omega}$,
also guarantees that $\ell\nu/L\ll1$ (see the discussion
around (\ref{eq:lambda_nu}) and section \ref{sec:consistency}),
there is a wide range of frequencies that satisfy the inequality (\ref{eq:omega_range}).
For frequencies within this range, (\ref{eq:ImG_smallOmega}) is a
universal result: The spectral function behaves as $\sim 
\mathrm{exp} \left(-\mathrm{const.} \cdot \hat{\omega}^{-1/(z-1)}\right)$,
and there are both constant and $\hat{\omega}$-dependent corrections
that can be computed order by order in perturbation theory. The naive
limit $\hat{\omega}\rightarrow0$ is non-universal, since higher derivative
corrections cannot be kept under control.

The procedure for calculating higher derivative corrections to the
spectral function outlined in this section can in principle be applied
to arbitrary corrections of the form (\ref{eq:U_WKB}). Note, however,
that since we generally expect an infinite number of such corrections,
going beyond leading order in $\ell/L$ may require expanding
(\ref{eq:S_WKB}) to the appropriate order.

Finally, let us comment on the sign of $\lambda_{i,j}$. In the analysis
above, we assumed that $\lambda_{i,j}<0$, so that the wavefunction
is always oscillating at the horizon, and no additional turning points
are introduced. If $\lambda_{i,j}$ is positive, the wavefunction
is tunneling in the deep IR, leading to another tunneling contribution
$S_{\mathrm{IR}}$ to the spectral function \eqref{eq:ImG_WKB}. At
large enough $\hat{\rho}$, the higher derivative corrections will
always dominate the potential, so $S_{\mathrm{IR}}$ does not have
a perturbative expansion in $\lambda_{i,j}$. This is simply a consequence
of the fact that the potential in the IR is always sensitive to all
of the (in principle infinitely many) coefficients that appear in the
series of higher derivative corrections, and we cannot solve the equation
of motion perturbatively in the IR. It therefore seems that one cannot
trust our analysis in the case of generic corrections with arbitrary
sign. However, one can circumvent this problem in the following way:
For any given $\hat{\omega}$, we can define a regulator surface at
$\hat{\rho}=\hat{\rho}_{h}(\hat{\omega})\gg\hat{\rho}_{0}$, such
that higher derivative corrections are still small at $\hat{\rho}_{h}$,
i.e. 
\begin{equation}
\lambda_{i,j}\hat{\omega}^{i}\hat{\rho}_{h}^{i+j/z}\ll\nu^{2},\qquad\lambda_{i,j}\hat{\omega}^{i}\hat{\rho}_{h}^{i+j/z}\ll\hat{\rho}_{h}^{2/z}.
\end{equation}
This guarantees that the wavefunction is still oscillating at $\hat{\rho}_{h}$,
even though eventually higher derivatives may cause the potential
to bend upwards again. Ignoring the (unknown) behavior of the wavefunction
in the deep IR, we only impose infalling boundary conditions at $\hat{\rho}_{h}$,
instead of $\hat{\rho}\rightarrow\infty$. The surface at $\hat{\rho}_{h}$
thus becomes an ``effective horizon'', where the wavefunction is
infalling: 
\begin{equation}
\psi(\hat{\rho}\rightarrow\hat{\rho}_{h})\approx ae^{i\Phi(\hat{\rho})}.
\end{equation}
Here $\Phi$ is an increasing function of $\hat{\rho}$. The retarded
Green's function can then be computed using the usual formula (\ref{eq:G=B/A}),
and the spectral function can be calculated approximately using the
WKB-formula \eqref{eq:ImG_WKB}. On a practical level, this regularization
prescription amounts to simply taking \eqref{eq:ImG_WKB}
for granted, and formally expanding the WKB integral $S$ in $\lambda_{i,j}$,
without worrying about the dynamics close to the horizon.

\section{Field theory models with $z=2$}
\label{sec:FT}

As we have demonstrated, a holographic computation of the spectral function yields
the universal low-frequency behavior $\chi\sim\exp(-\mathrm{const.}\cdot\hat\omega^{-1/(z-1)})$, provided
$\hat\omega$ is in the range (\ref{eq:omega_range}), where the higher derivative corrections
are controlled.  
From a field theory point of view, such an exponential behavior is not expected to arise
at any finite perturbative order, but can show up non-perturbatively.  This, of course, fits
the framework of non-relativistic holography, where the field theory dual is expected to
involve strong correlations. 

 In this section, we explore two field theoretic models exhibiting
$z=2$ Lifshitz scaling.  The first model is the quadratic band crossing model of
\cite{Sun2009}, and the second is the quantum Lifshitz model \cite{Ardonne2004}. Our strategy will be to identify phase-space regions with nonzero
decay rates for bosonic quasi-particles, which, according to the optical theorem, will contribute to the imaginary part of the corresponding bosonic Green's functions, and hence the spectral function.
In both models, we
confirm the presence of exponential suppression in the spectral function at small $\hat{\omega}$,
in agreement with the holographic computation.

\subsection{The quadratic band crossing model}

To set up the quadratic band crossing model, let us start with a massless Dirac theory in
$2+1$ dimensions, with action%
\footnote{Note that we use signature $(+,-,-)$ for the field theory.}
\begin{align}
S=\int d\vec{x} dt \left[\bar{\Psi}\left(i\gamma_0 \partial_0 - i \gamma_1 \partial_x 
-i \gamma_2 \partial_y  \right) \Psi - g \psi^\dagger_1\psi_2^\dagger\psi_2\psi_1\right] .
\label{eq:Dirac}
\end{align}
Here $\Psi=(\psi_1,\psi_2)^T$ is a two-component spinor and $\bar\Psi=\Psi^\dagger \gamma_0$.
The $2+1$ dimensional Dirac matrices are given by
\begin{align}
\gamma_0
=\left(\begin{matrix}
0 & -i\\
i & 0
\end{matrix}
\right), \;\;\;\;\;
\gamma_1
=\left(\begin{matrix}
0 & i\\
i & 0
\end{matrix}
\right), \;\;\;\;\;
\gamma_2
=\left(\begin{matrix}
-i & 0\\
0 & i
\end{matrix}
\right).
\end{align}
The interaction term in \eqref{eq:Dirac} is the only four-fermi term allowed for a two-component
spinor.  In the IR, this $\psi^4$ term is also the most relevant interaction term in the RG sense.
At the Gaussian fixed point, this theory is conformally invariant with dynamical critical exponent $z=1$. 
By setting the speed of light to unity,  the theory contains only one control parameter, which is the
interaction strength $g$.

The quadratic band crossing model generalizes the above to form a scaling invariant model with $z=2$
\cite{Sun2009}.  It does so by replacing the derivatives in the Dirac theory
\eqref{eq:Dirac} by the following operators:
\begin{align}
i\partial_0 &\to i\partial_0 + t_0 \nabla^2,\nn\\
i \partial_x &\to -t_1 (\partial_x^2-\partial_y^2),\nn\\
i \partial_y &\to -2 t_2 \partial_x \partial_y,
\end{align}
where $\nabla^2=\partial_x^2+\partial_y^2$ and $t_0$, $t_1$ and $t_2$ are real parameters.
After this substitution, we obtain
a model with $z=2$:
\begin{align}
S=\int d\vec{x} dt \left\{ \bar{\Psi}\left[\gamma_0 \left(i\partial_0+ t_0 \nabla^2 \right)
+\gamma_1 t_1\left (\partial_x^2-\partial_y^2\right)
 +2\gamma_2 t_2 \partial_x \partial_y\right]  \Psi  - g \psi^\dagger_1\psi_2^\dagger\psi_2\psi_1 \right\}.
\label{eq:action_general}
\end{align}
This action bears some resemblance with the original Dirac theory.
However, in direct contrast to the Dirac theory, 
whose action only contains first-order derivatives, this model has a first order time derivative and
second order spatial derivatives.  As a result, space and time have different scaling dimensions,
and it is straightforward to show that dimensionally $[t]=2[\vec{x}]$, corresponding to $z=2$
at the Gaussian fixed point. In condensed matter systems,  this model describes band touching
points with quadratic dispersions, which have been observed in bilayer graphene~(see for example the
review articles \cite{Nilsson2008,CastroNeto2009,Kotov2012}); a realization has been proposed in optical lattice systems, using high angular momentum orbitals~\cite{Sun2012}.

Generically, the action \eqref{eq:action_general} contains four control parameters:
$t_0$, $t_1$, $t_2$, and the interaction strength $g$. However, we can set one of the three
$t_i$'s to unity (say $t_2=1$) by rescaling. As shown in \cite{Sun2009} and detailed
in Appendix~\ref{app:qbc}, if we require SO(2) spatial rotational symmetry, then $t_1$ and
$t_2$ must coincide.  Furthermore, if a fermion particle-hole symmetry (i.e.\ charge conjugation) is enforced,
then $t_0$ must vanish.  Here we will focus on the case with $t_0=0$ and $t_1=t_2=1$,  which preserves
both the spatial rotational  and charge-conjugation symmetries.  In this case, the action reduces to
\begin{align}
S=\int d\vec{x} dt \left\{ \bar{\Psi}\left[\gamma_0 i\partial_0
+\gamma_1 \left (\partial_x^2-\partial_y^2\right)
 +2\gamma_2 \partial_x \partial_y\right]  \Psi  - g \psi^\dagger_1\psi_2^\dagger\psi_2\psi_1 \right\},
\end{align}
and as shown in Appendix~\ref{app:qbc}, the free dispersion relation for the two bands is given simply by
\begin{align}
\epsilon_{\pm}(\vec{k})=\pm k^2.
\label{eq:dispersion_special}
\end{align}
It is worth emphasizing that most of our conclusions remain valid 
as long as $|t_0|<|t_1|$ and $|t_0|<|t_2|$. As discussed in Appendix~\ref{app:qbc}, these
inequalities ensure that the model has both particles and holes in the weak coupling limit (small $g$).

\subsubsection{Renormalization group analysis}
At tree level, the $\psi^4$ term in the quadratic band crossing model is irrelevant (relevant) in the IR for systems above
(below) 2+1 dimensions.  In 2+1 dimensions, $g$ is marginal at the tree level. A one-loop RG
analysis indicates that a repulsive interaction ($g>0$) is marginally relevant at IR, 
while an attractive interaction $g<0$ is marginally irrelevant in the IR~\cite{Sun2009}. 
In the Dirac theory \eqref{eq:Dirac} on the other hand, the
$\psi^4$ term is irrelevant (relevant) in the IR for systems above (below) 1+1 dimensions.
In 1+1 dimensions, due to the special properties of  the 1+1 conformal group, the $\psi^4$ term
remains exactly marginal, before the system hits a Kosterlitz-Thouless transition.

\subsubsection{Boson correlation functions}
Although the model discussed above describes fermionic fields, bosonic modes can be constructed
from these fermionic degrees of freedom in the form of fermion bilinears. 
In the particle-hole channel, we can build four different fermion bilinears (boson modes)
\begin{align}
b_i= \bar{\Psi} \gamma_i \Psi,
\end{align}
with $i=0$, $1$, $2$, and $3$, and the fourth gamma matrix is given by
$\gamma_3=i \gamma_0\gamma_1\gamma_2$. Here $b_0$ is the fermion density
operator and the other three bosonic operators can be used as order parameters for various symmetry breaking phases (nematic or quantum 
anomalous Hall)~\cite{Sun2009}. At the Gaussian fixed point, these bosonic modes have $z=2$, which is inherited from the fermions.

\begin{figure}
\centering
        \subfigure[A one-loop contribution to the spectral function, corresponding to the decay of a boson into one particle-hole pair. ]{\includegraphics[width=.45\linewidth]{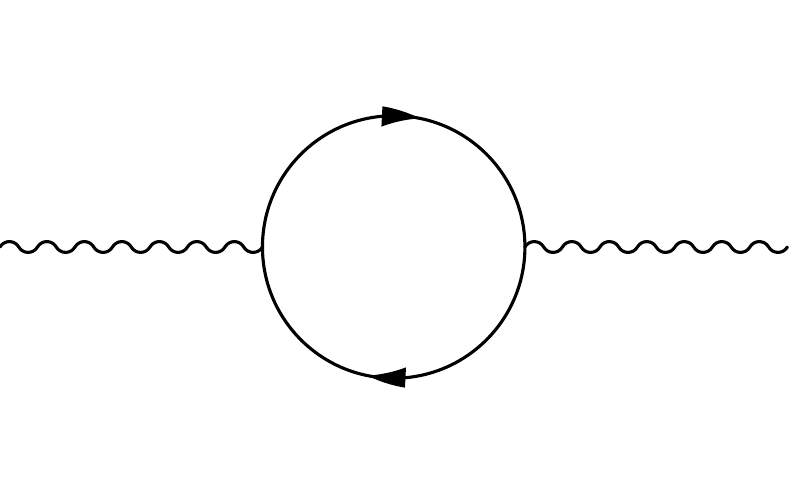}
                \label{fig:one-loop}}
                \hskip2em
        \subfigure[A five-loop contribution to the spectral function, corresponding to the decay into three particles and three holes.]{\includegraphics[width=.45\linewidth]{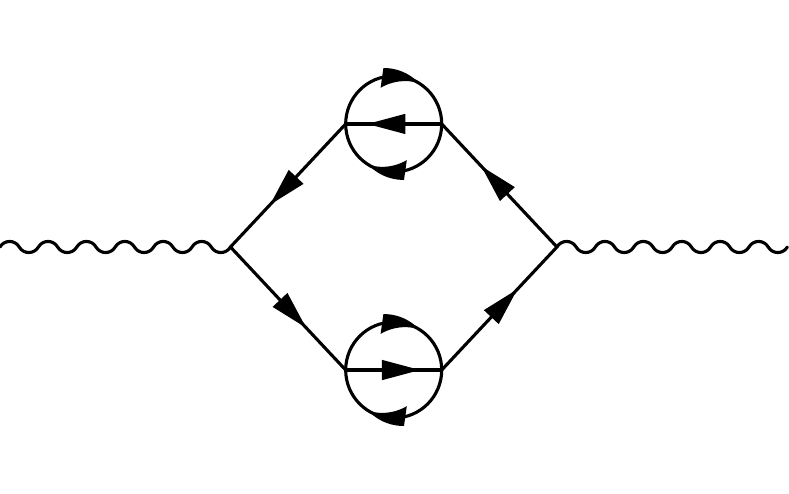}
                \label{fig:five-loop}}
\caption{Self-energy corrections for the boson modes. Here, solid lines represent fermionic propagators and wiggly lines are
boson propagators.}
\label{fig:diagrams}
\end{figure}

Additional bosonic modes can also be created in the particle-particle channel 
(e.g. $\psi_1^\dagger\psi_2^\dagger$), which are the order parameters for various superconducting states. In this section, 
we will only consider fermion bilinears in the particle-hole channel.  These bosons can decay into particle-hole pairs 
and are thus expected to have a finite lifetime. Via the optical theorem, the existence of such decay channels is equivalent to a non-zero imaginary part of the two-point function (and thus the spectral function), generated by self-energy diagrams such as those shown in  Fig.~\ref{fig:diagrams}.

Although it is challenging to analytically compute these diagrams,  it is straightforward to prove that for a boson with
momentum $\vec{k}$, the imaginary part of each self-energy diagram can only arise when the energy $\omega$ of the boson is 
larger than a certain threshold. For each diagram, this threshold can be determined using energy-momentum conservation. 

For example, the one loop diagram shown in Fig.~\ref{fig:one-loop} (the leading order correction) 
computes the scattering rate for a boson mode with energy $\omega$ and momentum $\vec{k}$ to decay into one particle 
with energy $\omega_p$ and momentum $\vec{k}_p$ and one hole with energy $\omega_h$ and momentum $\vec{k}_h$. 
Such a decay process can only take place when both the energy and momentum conservation laws are satisfied:
\begin{align}
\vec{k}&=\vec{k}_p-\vec{k}_h,\nn\\
\omega&=\omega_p-\omega_h=k_p^2+k_h^2\ge \frac{k^2}2.
\end{align}
Here we used the quadratic dispersion relation \eqref{eq:dispersion_special}.
For fixed $\vec{k}$,
the momentum conservation law enforces a relation between the momentum of the particle
$\vec{k}_p$ and that of the hole $\vec{k}_h$, i.e.\ $\vec{k}_p=\vec{k}+\vec{k}_h$.
With this constraint, the energy of the particle-hole excitation $k_p^2+k_h^2$ has a lower bound
of $k^2/2$ (which is reached when $\vec{k}_p=-\vec{k}_h=\vec{k}/2$). In other words,
the energy conservation law can only be satisfied when $\omega\ge k^2/2$. As a result,
for $\omega\ge k^2/2$, the boson can decay into a particle-hole pair, and thus have a finite
lifetime, while for $\omega<k^2/2$, decay is kinematically forbidden.  Thus,
at the one-loop level, $O(g^0)$, the imaginary part of the bosonic correlation function only arises for 
$\omega \ge k^2/2$. This energy range is known as the particle-hole continuum.

When the $\psi^4$-interaction term is taken into consideration, the bosonic modes can
decay through higher order processes (one example is shown in Fig.~\ref{fig:five-loop}).
For these higher order diagrams, the same analysis can be utilized. At order $O(g^{2n})$, 
the energy and momentum conservation laws imply that
\begin{align}
\vec{k}&=\sum_{i=1}^{n+1}\vec{k}_{p_i}-\sum_{i=1}^{n+1}\vec{k}_{h_i},\nn\\
\omega&=\sum_{i=1}^{n+1}\omega_{p_i}-\sum_{i=1}^{n+1}\omega_{h_i}
=\sum_{i=1}^{n+1} (k_{p_i}^2+k_{h_i}^2) \ge \frac{k^2}{2(n+1)}.
\end{align}
Here we consider the decay of a bosonic mode into $n+1$ particles and $n+1$ holes (see Fig.~\ref{fig:five-loop} for an example with $n=2$).
For fixed $\vec{k}$, momentum conservation enforces a constraint on the momenta of
the particles and holes.  Given this constraint, the energy is minimized when the momenta are
collinear, and the boson momentum $\vec{k}$ is equally distributed among the
particles and holes.  This results in a lower bound on the energy of $k^2/2(n+1)$.  Thus the decay
is kinematically forbidden unless $\omega\ge k^2/2(n+1)$. 

This analysis demonstrates that, up to order of $O(g^{2n})$, the imaginary part of the 
boson correlation function only arises when the energy of the boson is above a threshold,
$\omega\ge k^2/2(n+1)$.  Furthermore, this threshold goes down to zero for higher order
diagrams as $\sim 1/(n+1)$. Thus, if we sum the diagrammatic expansion to infinite order
($n\to\infty$), we expect that the boson correlation function can pick up a nonzero imaginary part for 
any $\omega>0$.

Finally, we are ready to extract the asymptotic form of the imaginary part of the self-energy correction at small
$\omega$. For $\omega\ll k^2$, the imaginary part  can only arise via high order process $O(g^{2n})$,
where $n\sim k^2/2\omega$. Therefore, we expect the imaginary part at energy $\omega$ and
momentum $\vec{k}$ to scale as $\sim g^{2n}\sim g^{k^2/\omega}$.  For sufficiently small $g$, this relation
implies that the imaginary part of the self-energy correction decays
to zero with the singular behavior $\sim e^{-\mathrm{const.}/\hat\omega}$, where
$\hat\omega=\omega/k^2$ is the dimensionless energy.
This matches the $z=2$ low frequency
behavior \eqref{eq:ImG_smallOmega} obtained holographically.

\subsubsection{Dirac theory revisited and systems with higher $z$}
We can repeat the kinematical analysis used above for similar models with arbitrary $z\ge 1$. In this case,
for $O(g^{2n})$,  the energy-momentum conservation law becomes
\begin{align}
\vec{k}&=\sum_{i=1}^{n+1}\vec{k}_{p_i}-\sum_{i=1}^{n+1}\vec{k}_{h_i},\nn
\\
\omega&=\sum_{i=1}^{n+1}\omega_{p_i}-\sum_{i=1}^{n+1}\omega_{h_i}=\sum_{i=1}^{n+1} (k_{p_i}^z+k_{h_i}^z)
\ge \frac{k^z}{(2n+2)^{z-1}}.
\label{eq:general_z}
\end{align}
For any $z>1$, the lower bound for having a nonzero imaginary part depends on $n$, and goes to zero as $n\rightarrow \infty$ (i.e. when considering higher and higher order diagrams). Similar to the discussion above, 
after summing over all the diagrams to infinite order, we find that at small $\omega$, the
imaginary part of the self-energy scales as 
\begin{align}
\mathrm{Im}\,\Pi \sim g^{(k^z/\omega)^{1/(z-1)}}.
\end{align} 
For small $g$, this indicates that $\mathrm{Im}\,\Pi$ decays to zero
as $e^{-\mathrm{const.}\cdot\hat\omega^{-1/(z-1)}}$, where now $\hat\omega=\omega/k^z$, in agreement with the holographic result
\eqref{eq:ImG_smallOmega}.  This suggests that the
exponential suppression of the spectral function is a generic property of Lifshitz models at $\omega\ll k^z$.

Note that for $z=1$, the Dirac theory is recovered, and the fate of the system is fundamentally different.
As can be seen by substituting $z=1$ into \eqref{eq:general_z}, the energy threshold
becomes independent of $n$. For any diagram, regardless of its order, the imaginary part arises only
for $\omega \ge k$. After summing over all diagrams (to infinite order), the same lower bound of
energy remains ($\omega \ge k$). As a result, for  $z=1$ the imaginary part of the correlation
function vanishes identically in a finite region $\omega \leq k$, which is in sharp contrast to the $z>1$ case. 
This conclusion is consistent with a symmetry analysis, which tells us that at $z=1$, the Lorentz
and conformal symmetries require the bosonic correlation function to be proportional to
$(-\omega+|\vec{k}|)^\alpha$, where $\alpha$ is some scaling exponent. For non-integer 
$\alpha$, $(-\omega+|\vec{k}|)^\alpha$ is real for $\omega<|\vec{k}|$, while the imaginary part arises for $\omega>|\vec{k}|$. For $z>1$, however, 
the absence of the Lorentz and conformal symmetries allows for very different types of behavior.

In summary, we find that models with $z>1$ and $z=1$ belong to fundamentally different universality classes. The case with $z=1$ (i.e. Dirac) has
been well understood with the help of conformal symmetry, which almost fully fixes the functional form of the correlation functions.
However, for $z>1$, the absence of conformal symmetry allows for richer structure in the correlation function. For arbitrary $z>1$,
we have presented an argument suggesting a characteristic exponential behavior $e^{-\textrm{const.}/\hat\omega^{1/(z-1)}}$ for the imaginary part of the self-energy correction
at low energy.

\subsubsection{Limitations of the analysis}
\label{sec:limit}
An exponential fall-off $\sim e^{-\textrm{const.}/\hat\omega^{1/(z-1)}}$ of the spectral function all the way down to $\omega\rightarrow 0$ would correspond to an essential singularity of the two-point function at the origin.
However, it is worth noting that there are two limitations of the analysis presented above. 
First, because we only considered the decay of bosonic modes into $n+1$ particle-hole pairs without taking into account the renormalization of the vertex 
function (i.e.\ the renormalization of the coupling constant $g$), the above analysis is not expected to give quantitatively accurate results in
the extremely low (or high) energy limit. This is because, as discussed above, in 2+1 dimensions, the coupling constant $g$ is marginally relevant or irrelevant (depending on the sign of $g$). For the IR or UV limit, the flow of $g$ cannot be ignored. However, because $g$ is only marginally relevant or irrelevant, 
the flow of $g$ is expected to be slow (i.e. logarithmic). Hence there may exist a range for $\omega$ (i.e. $\omega$ is small, but not too small) in which the RG flow of $g$ may be weak enough to be ignored, so that the analysis above can produce a reasonable estimate for the scaling behavior of $\mathrm{Im}\,\Pi$. 

Second, in the context of QFT,  the perturbation series in terms of Feynman diagrams is typically expected to be an asymptotic series. This means that our kinematical argument using loop diagrams only captures the behavior of the imaginary part of the self-energy correctly up to some finite order $O(g^{2N})$, where $N$ is large but finite. In particular, this implies that the scaling $\mathrm{Im}\, \Pi \sim g^{2n} \sim g^{1/\hat\omega^{1/(z-1)}}$ is only valid for $n \leq N$ and thus for $\hat \omega$ above some cutoff $\hat \omega_{\mathrm{\star}}(N)$.

Both of these points suggest that while the exponential suppression of the spectral function is a generic feature in a finite region where $\hat \omega$ is small, the behavior in the strict limit $\hat \omega\rightarrow 0$ is model-dependent. This is consistent with the observation in the gravity theory, where the would-be singular behavior of the two-point function may receive significant corrections at very small $\hat \omega$ from model-dependent higher derivative terms.

\subsection{The quantum Lifshitz model}
We now turn to the quantum Lifshitz model \cite{Ardonne2004}, which is a $z=2$ generalization of the Klein-Gordon theory. We start with the action
\begin{align}
S=\int d\vec{x} dt \left[\left(\partial_0\Phi \right)^2 -\left(\nabla^2\Phi \right)^2
-m \Phi^2 - g \Phi^4 \right].
\label{eq:Lifshitz}
\end{align}
Similar to the $\Phi^4$-model, positive (negative) $m$ corresponds to the disordered (ordered) phase respectively.
The quantum Lifshitz model focusses on the quantum critical point between these two phases (at $m=0$), 
at which the system is scaling invariant. At tree level, the dimensions of the various quantities are
\begin{align}
[\omega]=2, \qquad [k]=1, \qquad [\Phi]=\frac{d-2}{2},  \qquad [g]=6-d,
\end{align}
where $d$ is the number of spatial dimensions. The fact that $[\omega]=2[k]$ implies $z=2$.

In the non-interacting regime ($g=0$), the bosonic field $\Phi$ correspond to free bosons with quadratic dispersion relation
\begin{align}
\omega=\pm k^2,
\end{align}
and the (free) two-point correlation function is
\begin{align}
\avg{\Phi(k,\omega) \Phi(-k,-\omega)}=\frac{1}{\omega^2-k^4}.
\end{align}
\begin{figure}
\centering
        \subfigure[The two-loop sun-set 
diagram contribution to the spectral function, corresponding to the decay of one boson into three bosons.]{\includegraphics[width=.45\linewidth]{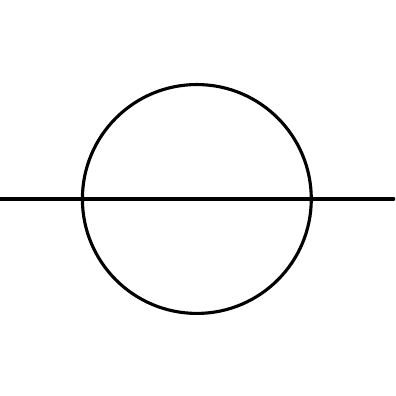}
                \label{fig:qone-loop}}
                \hskip2em
        \subfigure[A six-loop contribution to the spectral function, corresponding to the decay into seven bosons.]{\includegraphics[width=.45\linewidth]{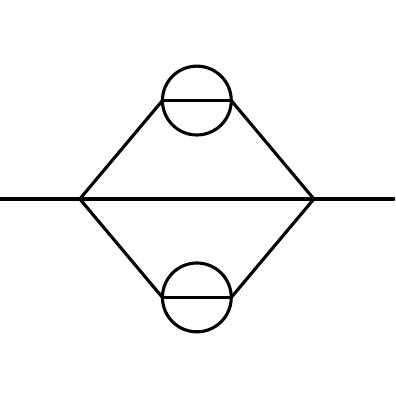}
                \label{fig:qfive-loop}}
\caption{Self-energy corrections in the quantum Lifshitz model. Here, solid lines represent bosonic propagators.}
\label{fig:qdiagrams}
\end{figure}
Similar to the previous model, we consider the decay of a bosonic mode with energy $\omega$ and momentum $\vec{k}$
into $2n+1$ bosonic modes (with energy $\omega_i$ and momentum $\vec{k}_i$ where $i=1, \ldots, 2n+1$) in the $2n$-th order
diagram $O(g^{2n})$. Energy and momentum conservation imply
\begin{align}
\vec{k}&=\sum_{i=1}^{2n+1} \vec{k}_i,\nn\\
\omega&=\sum_{i=1}^{2n+1} \omega_i=\sum_{i=1}^{2n+1} k_i^2\ge \frac{k^2}{2n+1}.
\end{align}
Once again, we find that the decay can only take place for $\omega$ above a threshold, which 
approaches zero as $n$ goes to infinity.

Using the same analysis, we see that the imaginary part of the correlation function is nonzero for any
finite $\omega$, and at small $\omega$ the imaginary part is $\sim g^{2n}$ with $n\sim k^2/2\omega$.
As a result, the imaginary part scales as $\sim g^{k^2/\omega}$, and we again recover the non-analytic
$z=2$ behavior $\sim e^{-\textrm{const.}/\hat\omega}$ as $\hat\omega\to0$.

\section{Discussion}
\label{sec:dis}

In our previous work, we found that the spectral function for a minimally coupled scalar in a Lifshitz background was  nonzero, but exponentially small, in the low-frequency regime $\omega \ll k^z$ \cite{Keeler:2014lia}.
The analysis presented here shows that this behavior is a robust holographic prediction for field theories with Lifshitz symmetry, in the absence of further constraining symmetries.
For the classes of higher derivative theories we study holographically, we generically find that the spectral function is suppressed in the low frequency region as $\chi\sim\exp(-\mathrm{const.}\cdot\hat\omega^{-1/(z-1)})$, so long as $\hat{\omega}\gg (\ell /L)^{z-1}$, where $\ell$ is the length scale at which higher derivatives become important. 

On the field theory side, the Lifshitz scaling symmetry is a priori not expected to lead to a universal 2-point function, and perturbative calculations do not reveal any similarities either between different field theories with Lifshitz symmetry, or with the holographic theory. 
However, we were able to show that in both of the field theory models considered here, a simple kinematical argument involving energy-momentum conservation and a resummation of loop diagrams reveals a similar exponential suppression as predicted by holography. Furthermore, this exponential suppression is expected for any field theory containing the following three key features:
The existence of particles and holes, an interaction that allows for decay channels, and a dispersion relation with $z>1$ scaling symmetry. Therefore, we expect our conclusion to be generic and applicable to a wide range of systems (with $z>1$) regardless of microscopic details, in agreement with the holographic prediction.

Although in both the holographic and the field theory calculation, the exponential suppression is a robust feature of the spectral function for small $\hat \omega$, the strict limit $\hat \omega \rightarrow 0$ is non-universal in both cases. In the holographic calculation, the model-dependence enters through higher derivative terms, which introduce corrections whose size can be quantified precisely (see equation (\ref{eq:omega_range})). However, the precise regime of validity of the field theory calculations is less clear. 
In both of the models considered here, the flow of the coupling constant $g$ can no longer be neglected when taking the exact limit $\hat{\omega} \rightarrow 0$. Instead of just being a simple exponential, the exact (nonperturbative) spectral function will therefore have a more complicated dependence on $\hat{\omega}$. Naively, one may expect a dependence of the form
\begin{equation}
\mathrm{Im}\,{\cal G}\sim g(\hat{\omega})^{\hat{\omega}^{-1/(z-1)}}.
\end{equation}
In 2+1 dimensions, the coupling $g$ is marginal, and we expect $g$ to depend only weakly on $\hat{\omega}$, so that the spectral function still shows an approximately exponential behavior. It would be interesting to further study the renormalization group flow of $g$ to make a
precise statement about the range of $\hat{\omega}$ for which this is the case. 
Along the same lines, in order to put a precise lower bound on $\hat \omega$, it would be important to account for the fact that the perturbative expansion is in fact only an asymptotic series (see the discussion in section \ref{sec:limit}).

In our field theory calculation, we found that $\mathrm{Im}\,{\cal G}\sim g^{1/ \hat{\omega}^{1/(z-1)}}$, so that exponential suppression in fact only arises for $g \ll 1$. It is important to note that this is not in contradiction to AdS/CFT being a weak-strong coupling duality. The strong coupling nature of the field theory does not necessarily mean that the parameter $g$ has to be chosen large, but rather that strong correlations (for example seen as long-range interactions) may emerge dynamically. This feature is familiar from the standard case of relativistic AdS/CFT, where it is not $g_{\mathrm{YM}}$ itself that is taken large, but rather the 't~Hooft coupling $g_{\mathrm{YM}}^2 N \gg 1$. In order to better understand the relation between strong/weak coupling on the field theory/gravity side in non-relativistic AdS/CFT, it would be desirable to develop a more precise version of the holographic dictionary for this case. 

Although we have chosen not to consider higher derivatives in the radial direction $\rho$ beyond second order, this is in
fact not a true limitation of the perturbative analysis.  Assuming we are only interested in solutions to the
higher derivative equation that are perturbatively connected to the lowest order
({\it i.e.}\ the two-derivative) equation, we may always eliminate higher derivatives by substituting in
the lower order equations.  Consider, for example, the addition of a fourth order term to the
Schr\"odinger-like equation (\ref{eq:Slehat})
\begin{equation}
-\psi''(\hat\rho)+\hat U(\hat\rho)\psi(\hat\rho)=\lambda\psi^{(4)}(\hat\rho).
\label{eq:rho4}
\end{equation}
We now rewrite this as $\psi''=U\psi-\lambda\psi^{(4)}$ and take two derivatives to obtain
$\psi^{(4)}=(U\psi)''-\lambda \psi^{(6)}$.  Substituting this in the right-hand side of
(\ref{eq:rho4}) and working only to linear order in $\lambda$ then reduces the equation to
second order
\begin{equation}
-\psi''(\hat\rho)+\hat U(\hat\rho)\psi(\hat\rho)-\lambda(\hat U(\hat\rho)\psi(\hat\rho))''=\mathcal O(\lambda^2).
\end{equation}
While this equation is no longer in manifest Schr\"odinger form, it can be so transformed if desired.
Thus our analysis is in fact applicable to this more general case as well.

As we discussed at the end of section \ref{sec:WKB}, the perturbative expansion of the spectral function in terms of higher derivative coefficients $\lambda_{i,j}$ strictly speaking only makes sense if these coefficients are chosen such that no additional turning points are introduced deep in the bulk. However, we argued that our formal perturbation series can still be used even in the case of higher derivatives with ``wrong'' sign, i.e. for the case where the effective potential bends upwards at large $\rho$. It would be interesting to determine if in a realistic theory, there are constraints on the signs of the coefficients $\lambda_{i,j}$, for example due to bulk causality or unitarity. It would also be interesting to study string theory embeddings of Lifshitz spacetimes, where the coefficients of higher derivative corrections can be determined exactly, and calculate the corrections to holographic correlation functions.

\section*{Acknowledgments}
This work was supported in part by the US Department of Energy under grant DE-SC0007859.  The work of CK is supported in part by the Danish Council for Independent
Research project ``New horizons in particle and condensed matter physics from black
holes''. KS is supported by the US National Science Foundation under grant NSF-PHY-1402971.

\appendix

\section{Perturbative expansion of the WKB integral}
\label{app:pewkb}
We would like to obtain an approximate expression for the WKB integral
(\ref{eq:S_WKB}) for a potential of the form
\begin{equation}
\hat{U}(\hat{\rho})=\hat{U}_{0}(\hat{\rho})+\mbox{\ensuremath{\delta\hat{U}}(\ensuremath{\hat{\rho}})},
\end{equation}
where 
\begin{equation}
\hat{U}_{0}=\frac{\nu^{2}}{\hat{\rho}^{2}}+\frac{1}{\hat{\rho}^{2-2/z}}-\hat{\omega}^{2},
\end{equation}
and $\mbox{\ensuremath{\delta\hat{U}}}$ represents a small correction
to the potential. To be precise, we assume that $\delta\hat{U}$ is
subdominant compared to the other terms in the potential for all $\hat{\rho}$
between the boundary and the classical turning point. To guarantee
this, it is sufficient to demand that 
\beq
\delta\hat{U}(\mbox{\ensuremath{\hat{\rho}}})\ll\hat{\omega}^{2}
\quad\mathrm{for}\quad0\leq\hat{\rho}\leq\hat{\rho}_{0}.
\eeq
We can then expand the turning point as follows:
\begin{equation}
\hat{\rho}_{0}(t)=\hat{\rho}_{0}^{(0)}\left(1+t+\cdots\right),
\end{equation}
where 
\begin{equation}
t\sim\frac{\delta\hat{U}(\hat{\rho}_{0}^{(0)})}{\hat{\omega}^{2}}\ll1,
\end{equation}
and $\hat{\rho}_{0}^{(0)}$ is the turning point of the unperturbed
potential, i.e.\ $\hat{U}_{0}(\hat{\rho}_{0}^{(0)})=0$. The relative
size of $\delta\hat{U}$ at the unperturbed turning point is what
controls the higher derivative expansion. The WKB integral (\ref{eq:S_WKB}) can
be written as $S=S^{(0)}+\delta S$, where
\begin{equation}
\delta S=S-S^{(0)}=\int_{\epsilon}^{\hat{\rho}_{0}(t)}d\hat{\rho}\sqrt{\hat{U}_{0}(\hat{\rho})+\mbox{\ensuremath{\delta\hat{U}}(\ensuremath{\hat{\rho}})}}-\int_{\epsilon}^{\hat{\rho}_{0}^{(0)}}d\hat{\rho}\sqrt{\hat{U}_{0}(\hat{\rho})}.\label{eq:delta_S_shift}
\end{equation}
One could attempt to simply expand the above expression formally in
$t$ and $\delta\hat{U}$, and it turns out that this does indeed
give the correct result (\ref{eq:delta_S}). However, this approach is problematic,
since $\hat{U}_{0}$ goes to zero at $\hat{\rho}_{0}^{(0)}$, and
thus the formal expansion parameter $\delta\hat{U}/\hat{U}_{0}$ blows
up at this location. The solution is to split up the integrals in
(\ref{eq:delta_S_shift}) in a way that the integrand always has a
well-defined expansion in terms of $\delta\hat{U}$. To do this, we
shift the first integral by rescaling $x\equiv\hat{\rho}\hat{\rho}_{0}^{(0)}/\hat{\rho}_{0}(t)$,
so that the upper bounds of both integrals are identical. We can then
combine both terms to obtain
\begin{equation}
\delta S\approx\int_{\frac{\epsilon}{1+t}}^{\epsilon}dx\sqrt{\frac{\nu^{2}}{x^{2}}+\frac{1+\frac{2}{z}t}{x^{2-2/z}}-(1+2t)+\delta\hat{U}\left(x\right)}+\int_{\epsilon}^{\hat{\rho}_{0}^{(0)}}dx\sqrt{\hat{U}_{0}(x)}\left[\sqrt{1+V(x)}-1\right],\label{eq:delta_S_split}
\end{equation}
where 
\begin{equation}
V(x)=\frac{\delta\hat{U}(x)+\frac{2}{z\hat{\rho}^{2-2/z}}t-2t}{\hat{U}_{0}(x)},
\end{equation}
and we expanded to linear order in $t$. The first term in (\ref{eq:delta_S_split})
is due to the shift of the lower bound of the first integral in (\ref{eq:delta_S_shift}).
Assuming that $\lim_{x\rightarrow0}x^{2}\delta\hat{U}(x)=0$, this
term evaluates to $\nu t$ after we send $\epsilon\rightarrow0$.
To compute the second integral, notice that although $\hat{U}_{0}(x)$
itself blows up at the upper bound, the ratio $V(x)$ remains finite
everywhere. Moreover, it is small by assumption, so we can expand
(\ref{eq:delta_S_split}) in terms of $V(x)$:
\begin{equation}
\delta S\approx\nu t+\int_{\epsilon}^{\hat{\rho}_{0}^{(0)}}dx\frac{\delta\hat{U}(x)+\frac{2}{z\hat{\rho}^{2-2/z}}t-2t}{2\sqrt{\hat{U}_{0}}}.
\end{equation}
The integral over the terms linear in $t$ exactly cancels the $\nu t$
term, and we arrive at the final result:
\begin{equation}
\delta S\approx\int_{\epsilon}^{\hat{\rho}_{0}^{(0)}}dx\frac{\delta\hat{U}(x)}{2\sqrt{\hat{U}_{0}}}.
\end{equation}
This is the first order correction to the WKB integral in the presence
of a perturbation $\delta\hat{U}$.


\section{The free quadratic band crossing theory}
\label{app:qbc}
In the absence of interactions ($g=0$), the action for the quadratic band crossing model,
shown in \eqref{eq:action_general}, describes two species of free fermions with
dispersion relations
\begin{align}
\epsilon_{\pm}(\vec{k})=\left(t_0 \pm \sqrt{\frac{(t_1^2+t_2^2)+(t_1^2-t_2^2) \cos 4 \theta_k}{2}}\right) k^2,
\label{eq:dispersion}
\end{align}
where $\epsilon_+(\vec{k})$ and $\epsilon_-(\vec{k})$  are the energies for the two species of fermions at momentum  $\vec{k}$ 
respectively. The angle $\theta_{k}$ is the azimuthal angle of $\vec{k}$. 

In contrast with the 2+1 dimensional Dirac theory, where the SO(2) spatial rotational symmetry arises automatically 
(even if the speed of light is different along $x$ and $y$, the rotational symmetry can be obtained by rescaling),
the $z=2$ model here in general only preserves a four-fold rotational symmetry (see the cosine term in the dispersion relation).
Continuous rotational symmetry is only recovered at $t_1=t_2$~\cite{Sun2009}.

For  $|t_0|<|t_1|$ and $|t_0|<|t_2|$, it is easy to realize that $\epsilon_+>0$ and $\epsilon_-<0$. Therefore, 
fermions with energy $\epsilon_+$ and $\epsilon_-$ are particles and holes respectively (or say particles and anti-particles). 
These particles and holes in general do not preserve the symmetry of charge conjugation, because $\epsilon_+\ne -\epsilon_-$, except when $t_0=0$.  Thus charge conjugation only becomes a symmetry at $t_0=0$.

If $|t_0|$ is larger than both $|t_1|$ and $|t_2|$, then $\epsilon_+$ and $\epsilon_-$ will have the same sign, so that they are both particles (or both holes). 
If $|t_1|\ne |t_2|$ and $|t_0|$ takes a value between $|t_1|$ and $|t_2|$, the system is anisotropic and along certain directions we have
a particle branch and a hole branch, but along certain other directions, we have two hole branches (or two particle branches).
There is one special case with $|t_0|=|t_1|=|t_2|$. Here, one of the fermions has zero energy for any momentum, and this is known as a {\it flat band}. 
For a flat band, higher order spatial derivatives become important and will in general lift the energy degeneracy. Because a flat band is 
typically associated with an infinite density of states, it is usually unstable (towards a certain symmetry breaking ground state) when interactions are introduced.

In the main text, we focus on the case with $t_0=0$ and $t_1=t_2=1$, but most conclusions remains valid
as long as  $|t_0|<|t_1|$ and $|t_0|<|t_2|$. For this special case, the dispersion relation reduces to 
$\epsilon_{\pm}(\vec{k})=\pm k^2$, as given in \eqref{eq:dispersion_special}.

\bibliography{Lifshitz}

\providecommand{\href}[2]{#2}\begingroup\raggedright\begin{thebibliography}{10}

\bibitem{Kachru:2008yh}
S.~Kachru, X.~Liu and M.~Mulligan, \emph{{Gravity Duals of Lifshitz-like Fixed
  Points}},
  \href{http://dx.doi.org/10.1103/PhysRevD.78.106005}{\emph{Phys.Rev.} {\bf
  D78} (2008) 106005}, [\href{http://arxiv.org/abs/0808.1725}{{\tt
  0808.1725}}].

\bibitem{Son:2008ye}
D.~Son, \emph{{Toward an AdS/cold atoms correspondence: A Geometric realization
  of the Schrodinger symmetry}},
  \href{http://dx.doi.org/10.1103/PhysRevD.78.046003}{\emph{Phys.Rev.} {\bf
  D78} (2008) 046003}, [\href{http://arxiv.org/abs/0804.3972}{{\tt
  0804.3972}}].

\bibitem{Balasubramanian:2008dm}
K.~Balasubramanian and J.~McGreevy, \emph{{Gravity duals for non-relativistic
  CFTs}},
  \href{http://dx.doi.org/10.1103/PhysRevLett.101.061601}{\emph{Phys.Rev.Lett.}
  {\bf 101} (2008) 061601}, [\href{http://arxiv.org/abs/0804.4053}{{\tt
  0804.4053}}].

\bibitem{Adams:2008wt}
A.~Adams, K.~Balasubramanian and J.~McGreevy, \emph{{Hot Spacetimes for Cold
  Atoms}}, \href{http://dx.doi.org/10.1088/1126-6708/2008/11/059}{\emph{JHEP}
  {\bf 0811} (2008) 059}, [\href{http://arxiv.org/abs/0807.1111}{{\tt
  0807.1111}}].

\bibitem{Keeler:2014lia}
C.~Keeler, G.~Knodel and J.~T. Liu, \emph{{Hidden horizons in non-relativistic
  AdS/CFT}}, \href{http://dx.doi.org/10.1007/JHEP08(2014)024}{\emph{JHEP} {\bf
  1408} (2014) 024}, [\href{http://arxiv.org/abs/1404.4877}{{\tt 1404.4877}}].

\bibitem{Leichenauer:2013kaa}
S.~Leichenauer and V.~Rosenhaus, \emph{{AdS black holes, the bulk-boundary
  dictionary, and smearing functions}},
  \href{http://dx.doi.org/10.1103/PhysRevD.88.026003}{\emph{Phys.Rev.} {\bf
  D88} (2013) 026003}, [\href{http://arxiv.org/abs/1304.6821}{{\tt
  1304.6821}}].

\bibitem{Keeler:2013msa}
C.~Keeler, G.~Knodel and J.~T. Liu, \emph{{What do non-relativistic CFTs tell
  us about Lifshitz spacetimes?}},
  \href{http://dx.doi.org/10.1007/JHEP01(2014)062}{\emph{JHEP} {\bf 1401}
  (2014) 062}, [\href{http://arxiv.org/abs/1308.5689}{{\tt 1308.5689}}].

\bibitem{Rey:2014dpa}
S.-J. Rey and V.~Rosenhaus, \emph{{Scanning Tunneling Macroscopy, Black Holes,
  and AdS/CFT Bulk Locality}},  \href{http://arxiv.org/abs/1403.3943}{{\tt
  1403.3943}}.

\bibitem{Hartong:2015zia}
J.~Hartong and N.~A. Obers, \emph{{Horava-Lifshitz Gravity From Dynamical
  Newton-Cartan Geometry}},  \href{http://arxiv.org/abs/1504.07461}{{\tt
  1504.07461}}.

\bibitem{Hofman:2014loa}
D.~M. Hofman and B.~Rollier, \emph{{Warped Conformal Field Theory as Lower Spin
  Gravity}},  \href{http://arxiv.org/abs/1411.0672}{{\tt 1411.0672}}.

\bibitem{Griffin:2012qx}
T.~Griffin, P.~Ho\^{r}ava and C.~M. Melby-Thompson, \emph{{Lifshitz Gravity for
  Lifshitz Holography}},
  \href{http://dx.doi.org/10.1103/PhysRevLett.110.081602}{\emph{Phys.Rev.Lett.}
  {\bf 110} (2013) 081602}, [\href{http://arxiv.org/abs/1211.4872}{{\tt
  1211.4872}}].

\bibitem{Sun2009}
K.~Sun, H.~Yao, E.~Fradkin and S.~Kivelson, \emph{Topological insulators and
  nematic phases from spontaneous symmetry breaking in 2d fermi systems with a
  quadratic band crossing},
  \href{http://dx.doi.org/10.1103/PhysRevLett.103.046811}{\emph{Phys. Rev.
  Lett.} {\bf 103} (Jul, 2009) 046811},
  [\href{http://arxiv.org/abs/0905.0907}{{\tt 0905.0907}}].

\bibitem{Ardonne2004}
E.~Ardonne, P.~Fendley and E.~Fradkin, \emph{Topological order and conformal
  quantum critical points}, {\emph{Annals Phys.} {\bf 310} (2004) 493--551},
  [\href{http://arxiv.org/abs/cond-mat/0311466}{{\tt cond-mat/0311466}}].

\bibitem{Son:2002sd}
D.~T. Son and A.~O. Starinets, \emph{{Minkowski space correlators in AdS / CFT
  correspondence: Recipe and applications}},
  \href{http://dx.doi.org/10.1088/1126-6708/2002/09/042}{\emph{JHEP} {\bf 0209}
  (2002) 042}, [\href{http://arxiv.org/abs/hep-th/0205051}{{\tt
  hep-th/0205051}}].

\bibitem{Policastro:2001yc}
G.~Policastro, D.~T. Son and A.~O. Starinets, \emph{{The Shear viscosity of
  strongly coupled N=4 supersymmetric Yang-Mills plasma}},
  \href{http://dx.doi.org/10.1103/PhysRevLett.87.081601}{\emph{Phys.Rev.Lett.}
  {\bf 87} (2001) 081601}, [\href{http://arxiv.org/abs/hep-th/0104066}{{\tt
  hep-th/0104066}}].

\bibitem{Policastro:2002se}
G.~Policastro, D.~T. Son and A.~O. Starinets, \emph{{From AdS / CFT
  correspondence to hydrodynamics}},
  \href{http://dx.doi.org/10.1088/1126-6708/2002/09/043}{\emph{JHEP} {\bf 0209}
  (2002) 043}, [\href{http://arxiv.org/abs/hep-th/0205052}{{\tt
  hep-th/0205052}}].

\bibitem{Policastro:2002tn}
G.~Policastro, D.~T. Son and A.~O. Starinets, \emph{{From AdS / CFT
  correspondence to hydrodynamics. 2. Sound waves}},
  \href{http://dx.doi.org/10.1088/1126-6708/2002/12/054}{\emph{JHEP} {\bf 0212}
  (2002) 054}, [\href{http://arxiv.org/abs/hep-th/0210220}{{\tt
  hep-th/0210220}}].

\bibitem{Herzog:2002fn}
C.~P. Herzog, \emph{{The Hydrodynamics of M theory}},
  \href{http://dx.doi.org/10.1088/1126-6708/2002/12/026}{\emph{JHEP} {\bf 0212}
  (2002) 026}, [\href{http://arxiv.org/abs/hep-th/0210126}{{\tt
  hep-th/0210126}}].

\bibitem{Herzog:2003ke}
C.~P. Herzog, \emph{{The Sound of M theory}},
  \href{http://dx.doi.org/10.1103/PhysRevD.68.024013}{\emph{Phys.Rev.} {\bf
  D68} (2003) 024013}, [\href{http://arxiv.org/abs/hep-th/0302086}{{\tt
  hep-th/0302086}}].

\bibitem{Kovtun:2004de}
P.~Kovtun, D.~T. Son and A.~O. Starinets, \emph{{Viscosity in strongly
  interacting quantum field theories from black hole physics}},
  \href{http://dx.doi.org/10.1103/PhysRevLett.94.111601}{\emph{Phys.Rev.Lett.}
  {\bf 94} (2005) 111601}, [\href{http://arxiv.org/abs/hep-th/0405231}{{\tt
  hep-th/0405231}}].

\bibitem{Buchel:2004di}
A.~Buchel, J.~T. Liu and A.~O. Starinets, \emph{{Coupling constant dependence
  of the shear viscosity in N=4 supersymmetric Yang-Mills theory}},
  \href{http://dx.doi.org/10.1016/j.nuclphysb.2004.11.055}{\emph{Nucl.Phys.}
  {\bf B707} (2005) 56--68}, [\href{http://arxiv.org/abs/hep-th/0406264}{{\tt
  hep-th/0406264}}].

\bibitem{Son:2007vk}
D.~T. Son and A.~O. Starinets, \emph{{Viscosity, Black Holes, and Quantum Field
  Theory}},
  \href{http://dx.doi.org/10.1146/annurev.nucl.57.090506.123120}{\emph{Ann.Rev.Nucl.Part.Sci.}
  {\bf 57} (2007) 95--118}, [\href{http://arxiv.org/abs/0704.0240}{{\tt
  0704.0240}}].

\bibitem{Nilsson2008}
J.~Nilsson, {A. H. Castro Neto}, F.~Guinea and N.~M.~R. Peres, \emph{Electronic
  properties of bilayer and multilayer graphene},
  \href{http://dx.doi.org/10.1103/PhysRevB.78.045405}{\emph{Phys. Rev. B} {\bf
  78} (2008) 045405}, [\href{http://arxiv.org/abs/0712.3259}{{\tt 0712.3259}}].

\bibitem{CastroNeto2009}
{A. H. Castro Neto}, F.~Guinea, N.~M.~R. Peres, K.~S. Novoselov and A.~K. Geim,
  \emph{The electronic properties of graphene},
  \href{http://dx.doi.org/10.1103/RevModPhys.81.109}{\emph{Rev. Mod. Phys.}
  {\bf 81} (2009) 109}, [\href{http://arxiv.org/abs/0709.1163}{{\tt
  0709.1163}}].

\bibitem{Kotov2012}
V.~Kotov, B.~Uchoa, V.~Pereira, F.~Guinea and A.~Castro~Neto,
  \emph{Electron-electron interactions in graphene: Current status and
  perspectives}, \href{http://dx.doi.org/10.1103/RevModPhys.84.1067}{\emph{Rev.
  Mod. Phys.} {\bf 84} (Jul, 2012) 1067--1125},
  [\href{http://arxiv.org/abs/1012.3484}{{\tt 1012.3484}}].

\bibitem{Sun2012}
K.~Sun, W.~V. Liu, A.~Hemmerich and S.~D. Sarma, \emph{Topological semimetal in
  a fermionic optical lattice},
  \href{http://dx.doi.org/doi:10.1038/nphys2134}{\emph{Nature Physics} {\bf 8}
  (2012) 67--70}, [\href{http://arxiv.org/abs/1011.4301}{{\tt 1011.4301}}].

\end{thebibliography}\endgroup
\bibliographystyle{JHEP}

\end{document}